\documentclass[fleqn,usenatbib]{mnras}
\usepackage[T1]{fontenc}

\DeclareRobustCommand{\VAN}[3]{#2}
\let\VANthebibliography\thebibliography
\def\thebibliography{\DeclareRobustCommand{\VAN}[3]{##3}\VANthebibliography}

\usepackage{graphicx}	% Including figure files
\usepackage{amsmath}	% Advanced maths commands
\usepackage{amssymb}	% Extra maths symbols

\usepackage{mathptmx}
\usepackage{txfonts}
\usepackage[english]{babel}
\title[Linking XLSSC 122 assembly to its members SFHs]{The XXL survey. XLIX. Linking the members star formation histories to the cluster mass assembly in the $\mathbf{z=1.98}$ galaxy cluster XLSSC 122
}

\author[A. Trudeau,  J. P. Willis, D. Rennehan et al.]{A. Trudeau$^{1}$\thanks{E-mail:\href{mailto:arianetrudeau@uvic.ca}{arianetrudeau@uvic.ca}}, J. P. Willis$^{1}$, D. Rennehan$^{1}$, R. E. A. Canning$^{2}$,  A. C. Carnall$^{3}$, B. Poggianti$^{4}$, E. Noordeh$^{5,6}$
\newauthor{and M. Pierre$^{7}$}
\\
$^{1}$Department of Physics \& Astronomy, University of Victoria, 3800 Finnerty Road, Victoria, British Columbia, V8W 2Y2, Canada\\
$^{2}$Institute of Cosmology and Gravitation, University of Portsmouth, Burnaby Road, Portsmouth, PO1 3FX, UK\\
$^{3}$Scottish Universities Physics Alliance, Institute for Astronomy, University of Edinburgh, Royal Observatory, Edinburgh EH9 3HJ, UK\\
$^{4}$INAF-Padova Astronomical Observatory, Vicolo dell'Osservatorio 5, I-35122 Padova, Italy\\
$^{5}$Department of Physics, Stanford University, 382 Via Pueblo Mall, Stanford, CA 94305-4060, USA\\
$^{6}$Kavli Institute for Particle Astrophysics and Cosmology, Stanford University, 452 Lomita Mall, Stanford, CA 94305-4085, USA\\
$^{7}$AIM, CEA, CNRS, Universit\'{e} Paris-Saclay, Universit\'{e} Paris Diderot, Sorbonne Paris Cit\'{e} 91191 Gif-sur-Yvette, France
}

\date{Accepted XXX. Received YYY; in original form ZZZ}

\pubyear{2022}

\begin{document}
\label{firstpage}
\pagerange{\pageref{firstpage}--\pageref{lastpage}}
\maketitle

 \defcitealias{pierre_xxl_2016}{XXL~Paper~I}
%   \defcitealias{2016A&A...592A...2P}{XXL~Paper~II}
%   \defcitealias{2016A&A...592A...3G}{XXL~Paper~III}
   \defcitealias{lieu_xxl_2016}{XXL~Paper~IV}
   \defcitealias{mantz_xxl_2014}{XXL~Paper~V}
%   \defcitealias{2016A&A...592A...5F}{XXL~Paper~VI}
%   \defcitealias{2016A&A...592A...6P}{XXL~Paper~VII}
%   \defcitealias{2016A&A...592A...7A}{XXL~Paper~VIII}
%   \defcitealias{2016A&A...592A...8B}{XXL~Paper~IX}
   \defcitealias{ziparo_xxl_2016}{XXL~Paper~X}
%   \defcitealias{2016A&A...592A..10S}{XXL~Paper~XI}
%   \defcitealias{2016A&A...592A..11K}{XXL~Paper~XII}
%   \defcitealias{2016A&A...592A..12E}{XXL~Paper~XIII}
%   \defcitealias{2016PASA...33....1L}{XXL~Paper~XIV}
   \defcitealias{lavoie_xxl_2016}{XXL~Paper~XV}
%%
%   \defcitealias{2018A&A...620A...1M}{XXL~Paper~XVI}
   \defcitealias{mantz_xxl_2018}{XXL~Paper~XVII}
%   \defcitealias{2018A&A...620A...3B}{XXL~Paper~XVIII}
%   \defcitealias{2018A&A...620A...4K}{XXL~Paper~XIX}
   \defcitealias{adami_xxl_2018}{XXL~Paper~XX.}
%   \defcitealias{2018A&A...620A...6M}{XXL~Paper~XXI}
%   \defcitealias{2018A&A...620A...7G}{XXL~Paper~XXII}
%   \defcitealias{2018A&A...620A...8F}{XXL~Paper~XXIII}
   \defcitealias{faccioli_xxl_2018}{XXL~Paper~XXIV}
%   \defcitealias{2018A&A...620A..10P}{XXL~Paper~XXV}
%   \defcitealias{2018A&A...620A..11C}{XXL~Paper~XXVI}
   \defcitealias{chiappetti_XXL_2018}{XXL~Paper~XXVII}
   \defcitealias{ricci_xxl_2018}{XXL~Paper~XXVIII}
%   \defcitealias{2018A&A...620A..14S}{XXL~Paper~XXIX}
%   \defcitealias{2018A&A...620A..15G}{XXL~Paper~XXX}
%   \defcitealias{2018A&A...620A..16B}{XXL~Paper~XXXI}
%   \defcitealias{2018A&A...620A..17P}{XXL~Paper~XXXII}
%   \defcitealias{2018A&A...620A..18L}{XXL~Paper~XXXIII}
%   \defcitealias{2018A&A...620A..19H}{XXL~Paper~XXXIV}
%   \defcitealias{2018A&A...620A..20K}{XXL~Paper~XXXV}
%

\begin{abstract}
The most massive protoclusters virialize to become clusters at $z\sim 2$, which is also a critical epoch for the evolution of their member galaxies. XLSSC 122 is a $z=1.98$ galaxy cluster with 37 spectroscopically confirmed members. We aim to characterize their star formation histories and to put them in the context of the cluster accretion history. We measure their photometry in 12 bands and create a PSF-matched catalogue of the cluster members. We employ BAGPIPES to fit star formation histories characterized by exponentially decreasing star-forming rates. Stellar masses, metal and dust contents are also treated as free parameters. The oldest stars in the red-sequence galaxies display a range of ages, from 0.5 Gyr to over $\sim$3 Gyrs. Characteristic times are between $\sim$0.1 and $\sim$0.3 Gyr, and the oldest members present the longest times. Using MultiDark Planck 2 dark matter simulations, we calculate the assembly of XLSSC 122-like haloes, weighted by the age posteriors of the oldest members. We found that 74\% of these haloes were less than 10\% assembled at the onset of star formation, declining to 67\% of haloes when such galaxies had formed 50\% of their z=1.98 stellar masses. When 90\% of their stellar masses were formed, 75\% of the haloes were less than 30\% assembled. The star formation histories of the red-sequence galaxies seem consistent with episodes of star formation with short characteristic times. Onset and cessation of star formation in the oldest galaxies are both likely to precede XLSSC 122 virialization.
%245 words
\end{abstract}

%%%%%%%%%%%%%%%%%%%%%%%%%%%%%%%%%%%%%%%%%%%%%%%%%%

%%%%%%%%%%%%%%%%% BODY OF PAPER %%%%%%%%%%%%%%%%%%

%\citep{mantz_xxl_2014}
%\citep[][also respectively referred as \citetalias{mantz_xxl_2014} and \citetalias{mantz_xxl_2018}]{mantz_xxl_2014,mantz_xxl_2018}
\section{Introduction}\label{sec_intro}

Clusters of galaxies are the most massive gravitationally bound structures in the Universe. At optical and near infrared wavelengths, they appear as overdensities of galaxies \citep{abell_distribution_1958}, hence the name. However, they are dark matter dominated objects ($\sim$85 per cent of the mass) with stars accounting for less than 3 per cent of their total masses \citep[e.g.][]{gonzalez_galaxy_2013,sanderson_baryon_2013,chiu_baryon_2016}. The bulk of their baryonic mass budgets is formed by a hot X-ray emitting gas called the intracluster medium.

Galaxy cluster progenitors are loosely bound, unvirialised overdensities of galaxies called protoclusters \citep{chiang_ancient_2013,muldrew_what_2015,lovell_characterising_2018}, which typically exhibit elevated star-forming rates (SFRs) and large amounts of cold gas \citep[e.g.][]{behroozi_average_2013,cucciati_discovery_2014,chiang_galaxy_2017,oteo_extreme_2018,miller_massive_2018}. These structures can be found as early as $z\sim 7-8$ \citep{ishigaki_very_2016,hu_lyman-_2021}, although more records exist for protoclusters at reshifts close to $z\sim 6.5$ \citep{franck_candidate_2016,calvi_mos_2019,chanchaiworawit_physical_2019,harikane_silverrush_2019,higuchi_silverrush_2019}. The number of these structures listed in the literature increases at lower redshifts \citep[e.g.][]{toshikawa_first_2014,wang_discovery_2016,jiang_giant_2018,kubo_planck_2019,kubo_massive_2021,shi_census_2019,shi_accelerated_2021,long_emergence_2020,calvi_probing_2021,kalita_ancient_2021}. 

Like clusters, protoclusters grow by merging with other structures \citep[e.g.][]{behroozi_average_2013,wu_rhapsody_2013,muldrew_what_2015,klypin_multidark_2016,werner_satellite_2022}. Theoretically, the difference between clusters and protoclusters is that clusters are virialized \citep[e.g.][]{chiang_ancient_2013,muldrew_what_2015,lovell_characterising_2018}. The epoch of the transition from protoclusters to clusters varies with their masses: the most massive overdensities collapse first, around $z\sim1.5-2$ \citep{chiang_ancient_2013,chiang_galaxy_2017,rennehan_rapid_2020}; the less massive ones might collapse as late as $z=0$. However, authors usually rely on observables to distinguish between protoclusters and clusters: a cluster should present an X-ray emitting intracluster medium and/or a red sequence \citep{papovich_spitzer-selected_2010,gobat_mature_2011,andreon_jkcs_2014,muldrew_what_2015}.

The red sequence is the line formed on a colour-magnitude diagram (CMD) by the reddest galaxies. Most of these galaxies are quenched, i.e. they ceased most of their star formation at least 0.5 Gyr before the epoch of observation. Quenching is an evolutionary process affected by stellar mass and environment: in every environment, massive galaxies tend to be more quenched \citep[e.g.][]{poggianti_evolution_2006,peng_mass_2010,peng_mass_2012,woo_dependence_2013,fossati_galaxy_2017,kawinwanichakij_effect_2017,jian_first_2018,lemaux_persistence_2019}. At $z\lesssim 1$, denser environments tend to increase the percentage of quenched galaxies for all masses \citep[e.g.][]{balogh_bimodal_2004,peng_mass_2010,peng_mass_2012,woo_dependence_2013,knobel_quenching_2015,balogh_evidence_2016,kawinwanichakij_effect_2017,jian_first_2018,cora_semi-analytic_2019,lemaux_persistence_2019,pintos-castro_evolution_2019}, although the effect seems more dramatic on low mass galaxies. At high redshifts, several studies suggest that the galaxy quenched fraction is already enhanced in dense environments \citep[e.g.][]{nantais_evidence_2017,lemaux_persistence_2019,strazzullo_galaxy_2019}, but the strength of this enhancement seems to depend on the host halo mass-scale \citep{cerulo_accelerated_2016,lemaux_persistence_2019}. Clusters of galaxies provide a gradient of dense environments, from the packed cores to the more scarcely populated outskirts \citep{balogh_origin_2000,poggianti_relation_2008,raichoor_galaxy_2012,aguerri_deep_2018,pintos-castro_evolution_2019,werner_satellite_2022} and are thus an ideal laboratory to understand how galaxy evolution is affected by environment at high redshifts.

Star formation in overdensities peaks at $z\sim 3$, earlier than in the field \citep[e.g.][]{behroozi_average_2013,chiang_galaxy_2017,muldrew_galaxy_2018}. This is also the epoch where the first signatures of quenching are expected \citep{poggianti_evolution_2006,lovell_characterising_2018,muldrew_galaxy_2018} a prediction supported by the recent observations of quenched galaxies in $z\sim3$ protoclusters \citep[e.g.][]{shi_census_2019,shi_accelerated_2021,kalita_ancient_2021,kubo_massive_2021}. The epoch between $z\sim 2$ and $z\sim 1.5$ represents a transition time with the coexistence of several quiescent and starbursting massive galaxies in clusters \citep[e.g.][]{strazzullo_galaxy_2013,andreon_jkcs_2014,fassbender_galaxy_2014,webb_extreme_2015,cooke_mature_2016,coogan_merger-driven_2018}. Later, at $z<1$, the evolution of the quenching fraction in clusters is mostly driven by intermediate and low-mass members \citep[e.g.][]{bower_precision_1992,poggianti_evolution_2006,raichoor_galaxy_2012,behroozi_average_2013,alberts_evolution_2014,aguerri_deep_2018,pintos-castro_evolution_2019}, the more massive galaxies being already quenched.

\begin{figure*}%[ht!]
\centering
%\begin{subfigure}{0.9\textwidth}
\centering
%\subfloat{
%\includegraphics[width=12cm]{XLSSC122_grid_age_chi2_black_triangles.eps}
%}
%%%\\
%%\begin{subfigure}{0.9\textwidth}
%\subfloat{
\includegraphics[width=12cm]{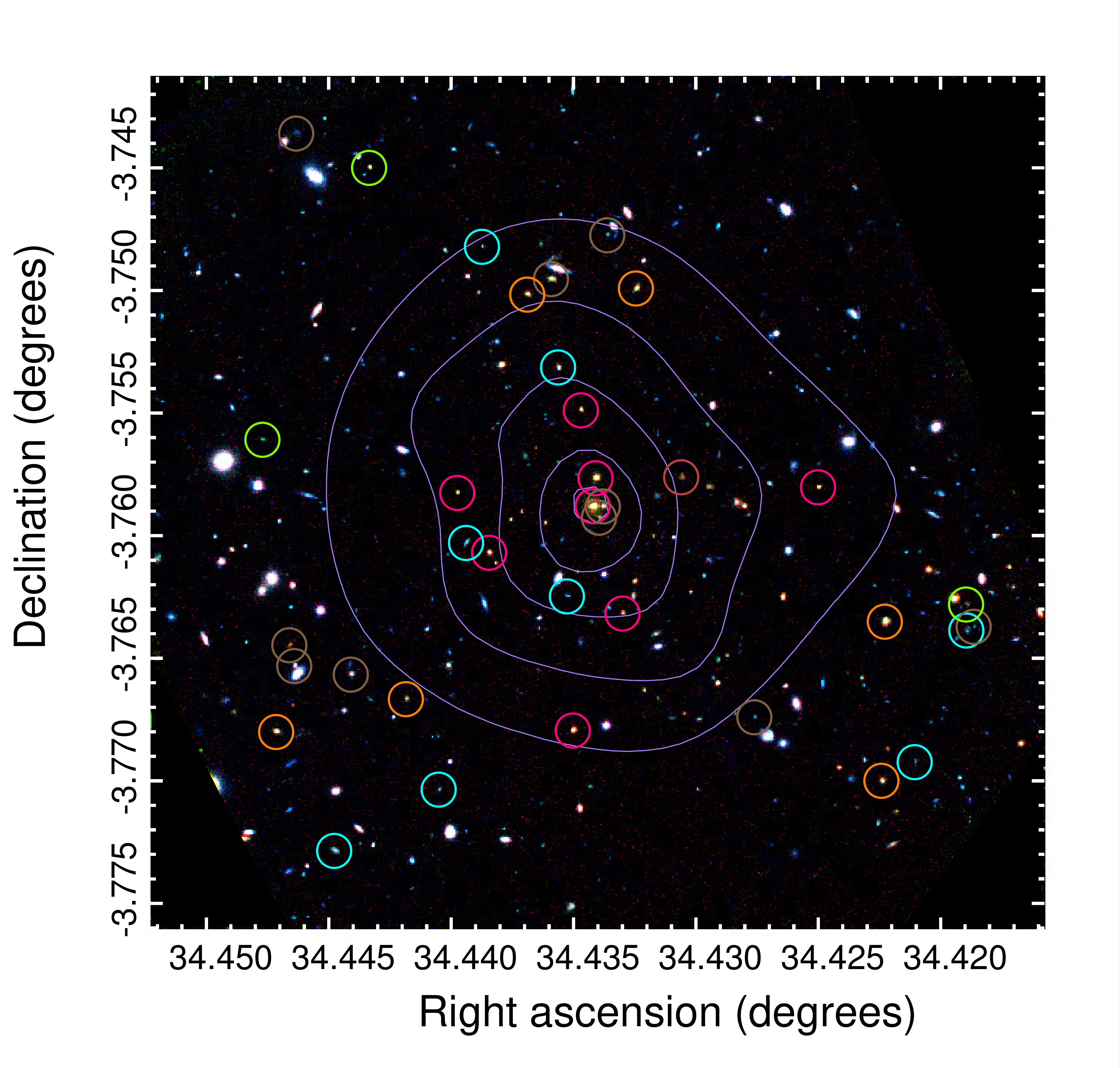}
\includegraphics[width=12cm]{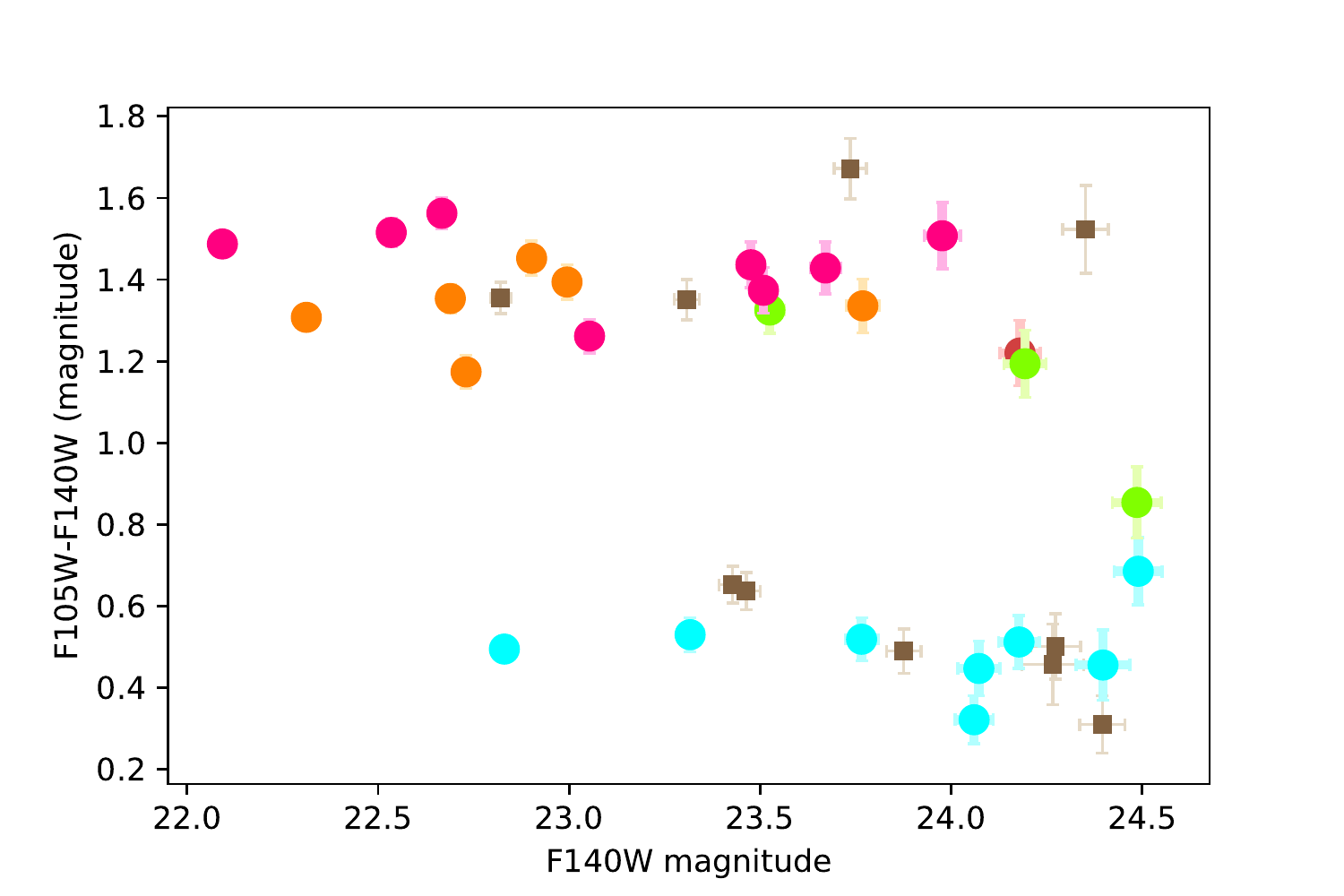}
%}
\caption{\textit{Top}: Colour image of XLSSC 122, made with three bands: \textit{F105W} (blue), \textit{F140W} (green) and $K_s$ (red). X-ray contours are drawn in violet. Spectroscopic members of the cluster are indicated by coloured circles. Colours refer to different ages (see Section \ref{ssec_age_tau}): pink for the oldest members, orange for the old ones, green for the young galaxies and cyan for the star-forming members. A dust-rich member is indicated in red. Members with poor fits (see the bronze members in Section \ref{ssec_grade}) are indicated in brown and not included in our analysis. \textit{Bottom}: CMD diagram on XLSSC 122 members, using the same colour code.}
\label{fig_XLSSC122_members}
\end{figure*}

XLSSC 122 is a $z=1.98$ galaxy cluster with 37 spectroscopically confirmed members \citep{willis_spectroscopic_2020} and a mass of $(6.3\pm 1.5)\times 10^{13}M_\odot$ \citep[][hereafter referred as \citetalias{mantz_xxl_2018}]{mantz_xxl_2018}. It is thus an interesting target to probe a critical epoch for both cluster assembly and members' evolution. The intracluster medium of XLSSC 122 is detected as an extended X-ray emission and also as a Sunayev-Zel'dovich decrement \citepalias[\citealt{mantz_xxl_2014}, also referred as \citetalias{mantz_xxl_2014} and][]{mantz_xxl_2018}. Recently, \citet{willis_spectroscopic_2020} found that XLSSC 122 has a prominent red sequence, with less luminous, less clustered blue cloud members. Figure \ref{fig_XLSSC122_members} summarizes these observations: the top panel shows a three-coloured image of XLSSC 122, with the members highlighted and the X-ray contours overlaid; the bottom panel presents the colour-magnitude diagram of those members.

In this paper, we use Spectral Energy Distribution (SED) modelling to reconstruct the star formation histories of XLSSC 122 members and compare them with the cluster assembly. Both seem to be linked \citep{poggianti_evolution_2006,muldrew_galaxy_2018,behroozi_universemachine_2019} but the mass-scale of the haloes in which the firsts passive galaxy were quenched is unclear. Section \ref{sec_catalogue} presents the building of the multiwavelength catalogue used to model the member SEDs. Section \ref{sec_results} presents the detail of this modelling and its main results, which are discussed and compared with the cluster assembly in Section \ref{sec_discussion}. A summary of our main findings is given in Section \ref{sec_summary}. We assume $H_0=70~\mathrm{km~s^{-1} Mpc^{-1}}$ with $\Omega_m=0.3$ and $\Omega_\Lambda=0.7$. Hence, at $z=1.98$ the age of the Universe is 3.26 Gyrs and one arcsec corresponds to 8.38 kpc.

\section{Construction of a multiwavelength catalogue}\label{sec_catalogue}

Our catalogue (available in the supplementary material online) contains 12 bands, which includes two \textit{Hubble Space Telescope} (\textit{HST}) images (\textit{F105W} and \textit{F140W}) and seven images coming from publicly available surveys. We direct the reader toward \citet{willis_spectroscopic_2020} for a description of the data reduction process for the \textit{HST} Wide-Field Camera 3 images. \textit{Y}, \textit{J} and $K_s$-band images were taken by the High Acuity Wide field K-band Imager (HAWK-I) mounted on the Very Large Telescope \citep[VLT;][]{pirard_hawk-i_2004,casali_hawk-i_2006,kissler-patig_hawk-i_2008,siebenmorgen_science_2011}. We also use observations from the Canada-France-Hawaii Telescope Legacy Survey \citep[CFHTLS;][]{gwyn_canada-france-hawaii_2012}, in the \textit{u}, \textit{g}, \textit{r} and first generation \textit{i} band. CFHTLS \textit{z}-band image is heavily affected by fringing; we replaced it with \textit{z}-band observations from the Hyper Suprime Camera Subaru Strategic Program \citep[HSC-SSP;][]{aihara_hyper_2018,aihara_first_2018}. We do not use other HSC-SSP observations as, surprisingly, an estimate of the depths based on the \textsc{SExtractor} noises showed that the corresponding CFHTLS observations are deeper. We completed the catalogue with \textit{I1} and \textit{I2} images taken by the Infrared Array Camera (IRAC), as part of the \textit{Spitzer} Wide Infrared Extragalatic Survey (SWIRE). 

Our catalogue consists of 37 entries, one for each spectroscopic member of XLSSC 122. These members were selected to be brighter than a magnitude of 25.5, measured within a 0.8 arcsecond-wide aperture in the \textit{F140W} filter \citep{willis_spectroscopic_2020}. At $z=1.98$, the \textit{F140W} filter probe the SED beyond the 4000 \AA~break, making the sample selection relatively independant of the galaxy star-formation history. We note however that the \textit{F140W} magnitude might be influenced by the presence of strong emission or absorption lines such as [OIII], H $\beta$ or H $\gamma$, resulting in a weak bias toward star-forming galaxies in the faintest part of our sample.

\subsection{Source extraction and aperture selection}\label{ssec_apertures}
Source Extraction was performed with \textsc{SExtractor} version 2.5.0 \citep{bertin_sextractor_1996}. \textsc{SExtractor} detects the \textit{F140W} sources using a pixel-based inverse variance weighting (IVM) and weights the photometry with the root mean square (RMS) variation per pixel. To facilitate source matching, \textit{F105W} photometry is computed in dual image mode, using \textit{F140W} as the detection image. These bands probe the SED before and after the 4000 \AA~break respectively. Thus, red-sequence galaxies are usually brighter in the \textit{F140W} image, which motivates our choice to select it as our reference image for aperture corrections.

The \textit{F105W} image is convolved with a Moffat kernel \citep{moffat_theoretical_1969} to match its resolution with the \textit{F140W}-band resolution. A suitable aperture for the \textit{HST} images is selected by locating the confirmed members of XLSSC 122 in the \textit{F140W} image. Their average growth curve is computed using 10 apertures between 0.18 and 2.28 arcsec in diameter. %: 0.18, 0.36, 0.54, 0.78, 1.02, 1.20, 1.38, 1.68, 1.98 and 2.28 arcsec in diameters. 
We apply strict contamination criteria: any members with neighbours brighter than its magnitude minus 3 (i.e. brighter than 6.4 per cent of its flux density) and closer than twice the full width at half maximum (FWHM) of the contaminating source is not included in the computation. We also did not include the Brightest Cluster Galaxy (BCG) and the three members closest to it in the aperture calculation, as they might be contaminated by intracluster light. However, we keep the BCG and ID 657 in the silver sample of our SED fitting results (see Section \ref{ssec_grade} and following). We aim to include about 80 per cent of the flux density within the chosen aperture. Thus, for the \textit{HST} images, we selected the 1.02 arcsec aperture as the most suitable one (see Table \ref{table_best_apertures}).

We noticed a systematic offset between the \textit{HST} images and their ground-based or \textit{Spitzer} counterpart. Thus, we compute an astrometric correction of the first order using 200 sources, which were selected to possess \textit{F140W} magnitudes similar to those of XLSSC 122 members. Each band photometry is computed in dual image mode, using the $K_s$ band as the detection image. This choice was motived by two considerations. First, \textsc{SExtractor} dual image mode can be used only on images with the same pixel size, which means that some bands must be resampled. Compared to \textit{HST}, HAWK-I pixel size is closer to the original CFHT, HSC and IRAC pixel sizes. Second, sources can be reliably matched between the \textit{F140W} and $K_s$ band: 90 per cent of the 200 sources used for astrometry matching have separations inferior to 0.24 arcsec between these two bands.

The aperture sizes of the ground-based images are chosen as described above, but with test apertures between 0.64 and 5.11 arcsec, which are better adapted to the resolutions of these bands. Results are shown in Table \ref{table_best_apertures}. However, due to the lower resolution of \textit{Spitzer} IRAC, we are unable to find a sufficient number of isolated XLSSC 122 members to compute an average member growth curve for \textit{I1} and \textit{I2}. We thus assume that they are unresolved and choose apertures containing approximately 80 per cent of the flux density of a point source. The IRAC data are also subject to colour-dependent flat-field errors. We account for them by increasing the IRAC flux density errors by 10 per cent of the flux density values, added in quadrature.

\begin{table}
\caption{Summary of the sizes and properties of the applied apertures. The third column presents the percentage of the average flux densities within the apertures.} 
\label{table_best_apertures}
\centering
\setlength{\tabcolsep}{2pt}
%%\tablenum{1}
\begin{tabular}{c c c c}
\hline
Band & Aperture diameter & Flux within & 5$\sigma$ detection\\
 & (arcsec) & (\%) & (mag)\\
\hline

\textit{u} & 2.66 & $82\pm 5$ & 25.9\\ % 25.94 0.8144580763764967 0.05057658630853357
\textit{g} & 2.45 & $83\pm 4$ & 26.1\\ %26.12 0.8326200801415292 0.03514760165953872
\textit{r} & 1.81& $77\pm 4$ & 25.9\\ %25.91 0.7657909675587281 0.03851494813770733
\textit{i} & 2.13 & $78\pm 3$ & 25.6\\ %25.62 0.8278901882530778 0.0354060618365692
\textit{z} & 1.81 & $77\pm 5$ & 25.0\\ %25.04 0.7801603791747631 0.03615326264695923
\textit{Y} & 1.81 & $85\pm 6$ & 24.8\\ %24.80 0.7823711040390539 0.04993348156108329
\textit{F105W}$^a$ & 1.02 & - & 26.87\\ %26.87 0.8303131936604726 0.009509201291243247
\textit{J} & 2.13 & $81\pm 4$ & 24.2\\ %24.20 0.8103672993027089 0.029154459429232345
\textit{F140W} & 1.02 & $77.8\pm 0.8$ & 25.96\\ %25.96 0.8303131936604726 0.009509201291243247
$K_s$ & 1.38 & $80\pm 2$ & 24.3\\ %24.26 0.7807765436505127 0.01350802603979102
\textit{I1}$^b$ & 3.73 & $78\pm 3$ & 23.2\\ %23.22 0.8230354867685284 0.035222394055447234
\textit{I2}$^b$ & 4.26 & $82\pm 3$ & 22.5\\ %22.53 0.827918665018535 0.04064301238472156
\hline
\multicolumn{4}{p{7.5cm}}{$^a$ Since they have similar PSFs, we assumed that the best aperture for the convolved \textit{F105W} image is equivalent to the best aperture for \textit{F140W}.}\\
\multicolumn{4}{p{7.5cm}}{$^b$ Apertures and percentages are based on the point source growth curves.}\\        
\end{tabular}
\end{table}

\subsection{Correction factors}\label{ssec_corr_facts}

To account for the difference in resolution between the \textit{HST} images and the other data, we apply aperture correction factors. We base our computations on the \citet{newman_can_2012} method: we convolve our \textit{F140W} image with different Moffat kernels. Then, we select the convolved images possessing the closest point spread function (PSF) to the ground-based and IRAC images by comparing the average growth curves of their point sources. 

The aperture correction factors are computed object-by-object, by comparing the galaxy flux density in the original \textit{F140W} image to its flux density in the convolved image. Thus, for band X, the correction factor $a_X$ and the corrected flux density $F_{X,corr}$ are given by:

%\begin{equation}
%a^2=b^2+c^2
%\end{equation}

\begin{equation}\label{eq_corr_fact}
a_X=\frac{F_{140W}}{F_{140W,X}}
\end{equation}

\noindent and,

\begin{equation}\label{eq_corr_flux}
F_{X,corr}=a_X F_X,
\end{equation}

\noindent where $F_{140W}$ is the flux density of the considered source in the \textit{F140W} image, measured in a 1.02 arcsec aperture, and $F_{140W,X}$ is the flux density in the convolved image, measured within the aperture selected for band X. $F_X$ is the flux density measured in band X. Since we are interested in 37 galaxies only, all within a single image, we deemed it preferable to compute correction factors for every cluster member rather than one per band as \citet{newman_can_2012} did. Furthermore, correction factors are more spread per band than they are per object.
To account for possible small differences between the convolved and ground-based image PSFs, we add in quadrature 5 per cent of the correction factor value to its errors. 

\section{Results}\label{sec_results}

\subsection{SED modelling}\label{ssec_fitting}

We perform the SED modelling with the python-based package Bayesian Analysis of Galaxies for Physical Inference and Parameter EStimation (\textsc{Bagpipes}) version 0.8.4 \citep{carnall_inferring_2018}. \textsc{Bagpipes} is a Bayesian SED modelling and fitting code, that generates SED models based on \citet{bruzual_stellar_2003} stellar population models and explores the parameters space with MultiNest \citep{feroz_multimodal_2008,feroz_multinest_2009,feroz_importance_2019,buchner_statistical_2014} a nestled sampling algorithm. We use uniform priors. The merit function quantifying the match between a model and the photometric data is evaluated by the following likelihood \citep{carnall_inferring_2018}:

\begin{equation}\label{eq_likelihood}
\mathrm{ln} (\mathcal{L})=-0.5\sum_i \mathrm{ln} \left(2\pi\sigma_i^2\right) -0.5 \sum_i \frac{\left(f_i-f_{m,i}\right)^2}{\sigma_i^2},
\end{equation}

\noindent
where $f_i$ are the observed flux densities (in $\mathrm{\mu}$Jy), $\sigma_i$ their errors, and $f_{m,i}$ are the flux densities predicted by the tested model.

We assume a simple star formation history, described by five free parameters. Two describe an exponentially decreasing star-forming rate: the age of the oldest stars and the characteristic time ($\tau$). The other free parameters control the dust extinction ($A_v$), following a \citet{calzetti_dust_2000} extinction curve, the metallicity ($Z$) and the logarithm of the total stellar mass formed, expressed in solar units. Prior intervals displayed in Table \ref{table_priors} are uniformly weighted. Redshifts are set to the values measured by \cite{willis_spectroscopic_2020} and each MultiNest run is initiated with a thousand live points.

\begin{table}
\caption{List of the parameter priors, expressed in term of minimum and maximum allowed values.}
\label{table_priors}
\centering
%\setlength{\tabcolsep}{2pt}
%%\tablenum{1}
\begin{tabular}{c c c c }
\hline
Parameter & Unit & Minimum & Maximum\\
\hline
Age & Gyr & 0.01 & 3.26$^a$\\
$\tau$ & Gyr & 0.01 & 3.26$^a$\\
$A_v$ & mag & 0 & 2.5\\
$Z$ & $Z_\odot$ & 0 & 5\\
Mass & $\mathrm{log} \left( \frac{M}{M_\odot} \right)$ & 9 & 13\\
\hline
\multicolumn{4}{l}{$^a$ Age of the Universe at z=1.98.}\\        
\end{tabular}
\end{table}

\subsection{Assessing the fit quality}\label{ssec_grade}

We assess the quality of the resulting fits with the following criteria:

\begin{enumerate}
\item The number of degrees of freedom in the fit must be one or more. This effectively means that a member must be detected in at least 6 bands. 
\item There should be flux measurements in both \textit{HST} bands (\textit{F105W} and \textit{F140W}). These two bands put the furthest constrains on the fit, as they are the deepest and probe the SED before and after the 4000 \AA~break.
\item The most likely SED model must have a reduced $\chi^2$ ($\chi^2_\nu$) inferior to 6. For a fit with seven degrees of freedom (i.e. with detections in every band), $\chi^2_\nu=6$ corresponds to a difference of $5\sigma$ between the fit and the data points.
\end{enumerate}

These criteria are met by 26 members, which are highlighted by coloured circles and dots in Figure \ref{fig_XLSSC122_members}. The 11 other members, called \lq bronze members\rq , are indicated in brown and are not included in our analysis. 

For those 26 members, we define another criterion, meant to distinguish between those with well-behaved photometric data (gold members) and those where contamination by nearby sources might make the photometry less reliable:
\begin{enumerate}
\setcounter{enumi}{3}
\item When assembling the aperture corrected photometry of members, we flagged photometric data that might be blended with another object, or significantly polluted by the light of a neighbouring detection. We require a good fit to have no photometric data points flagged, with the exception of the two IRAC data points, which tend to be systematically contaminated because of their resolution.
\end{enumerate}

\subsection{Age and characteristic time}\label{ssec_age_tau}

We find that the ages of the oldest stars in XLSSC 122 form a continuum: the peaks of the age distributions (i.e. the modal ages) vary from less than 0.2 Gyr to close to the age of the Universe (3.26 Gyrs at z=1.98). Red-sequence members all have short characteristic times ($\lesssim$ 0.3 Gyr), in contrast to poorly constrained $\tau$ displayed by most of the blue cloud members. We define four main categories of members, based on the medians of the age and characteristic time posteriors:

\begin{enumerate}
\item Very old members: display median ages above 1.75 Gyrs. %Throughout the figures of this paper, very old members are displayed in pink.
\item Old members: display median ages between 0.75 and 1.75 Gyrs. %In figures, they are shown in orange.
\item Young members: display ages younger than 0.75 Gyr, with median $\tau$ shorter than 1 Gyr. %They are displayed in green.
\item Star-forming members: display ages younger than 0.75 Gyr, with median $\tau$ greater or equal to 1 Gyr. %In figures, they appear in cyan.
\end{enumerate}

\begin{figure}
\centering
\includegraphics[width=8cm]{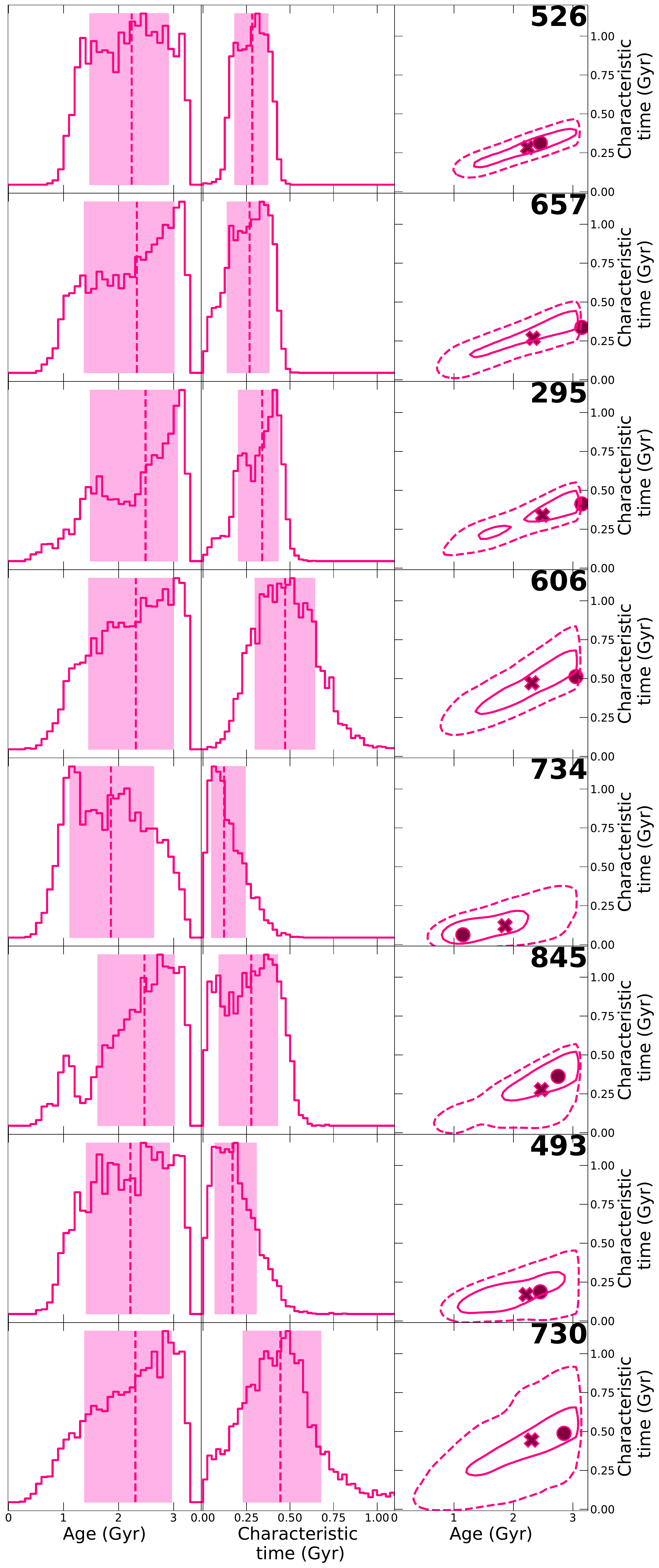}

\caption{\textit{Left}: Posterior distributions of the age of the oldest stars for every member classified as very old. The dashed lines are the distribution medians and the shaded regions correspond to the intervals between the 16th and 84th percentiles. \textit{Centre}: Posterior distributions for the characteristic time. \textit{Right}: Smoothed 1 and 2$\sigma$ contours for the degeneracy between age and $\tau$. The distributions medians and modes are represented by Xs and dots respectively. Each member ID is shown on the top right corner of its degeneracy plot.}% A representative very old member, ID 526, is highlighted with thicker lines and font.}
\label{fig_very_old_members}
\end{figure}

\begin{figure}%[h!]
\centering
\includegraphics[width=8cm]{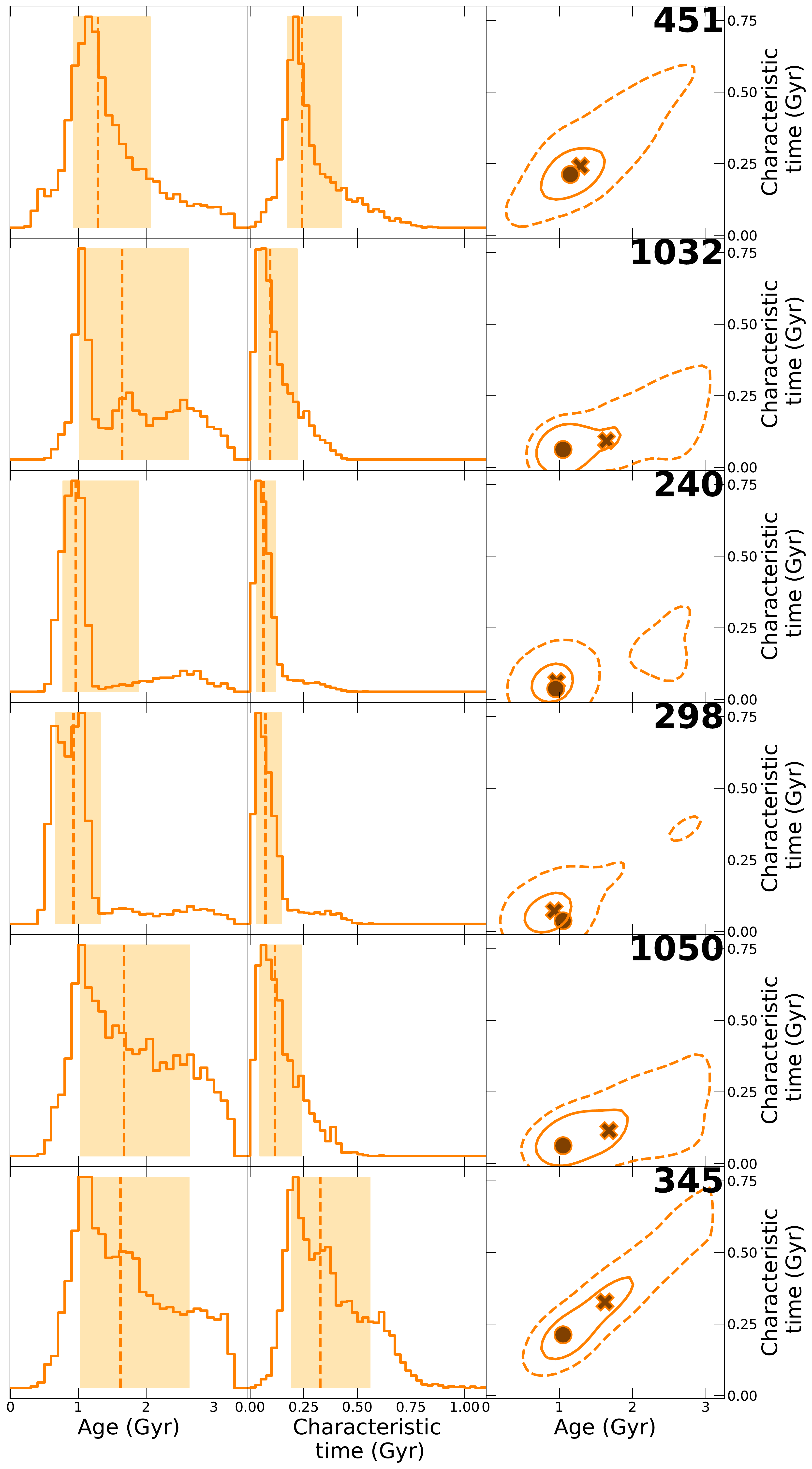}

\caption{Posterior distributions of the ages and characteristic times of the old members, along with their degeneracies. Lines and contours definition are given in Figure \ref{fig_very_old_members}.}
\label{fig_old_members}
\end{figure}

\begin{figure}%[h!]
\centering
\includegraphics[width=8cm]{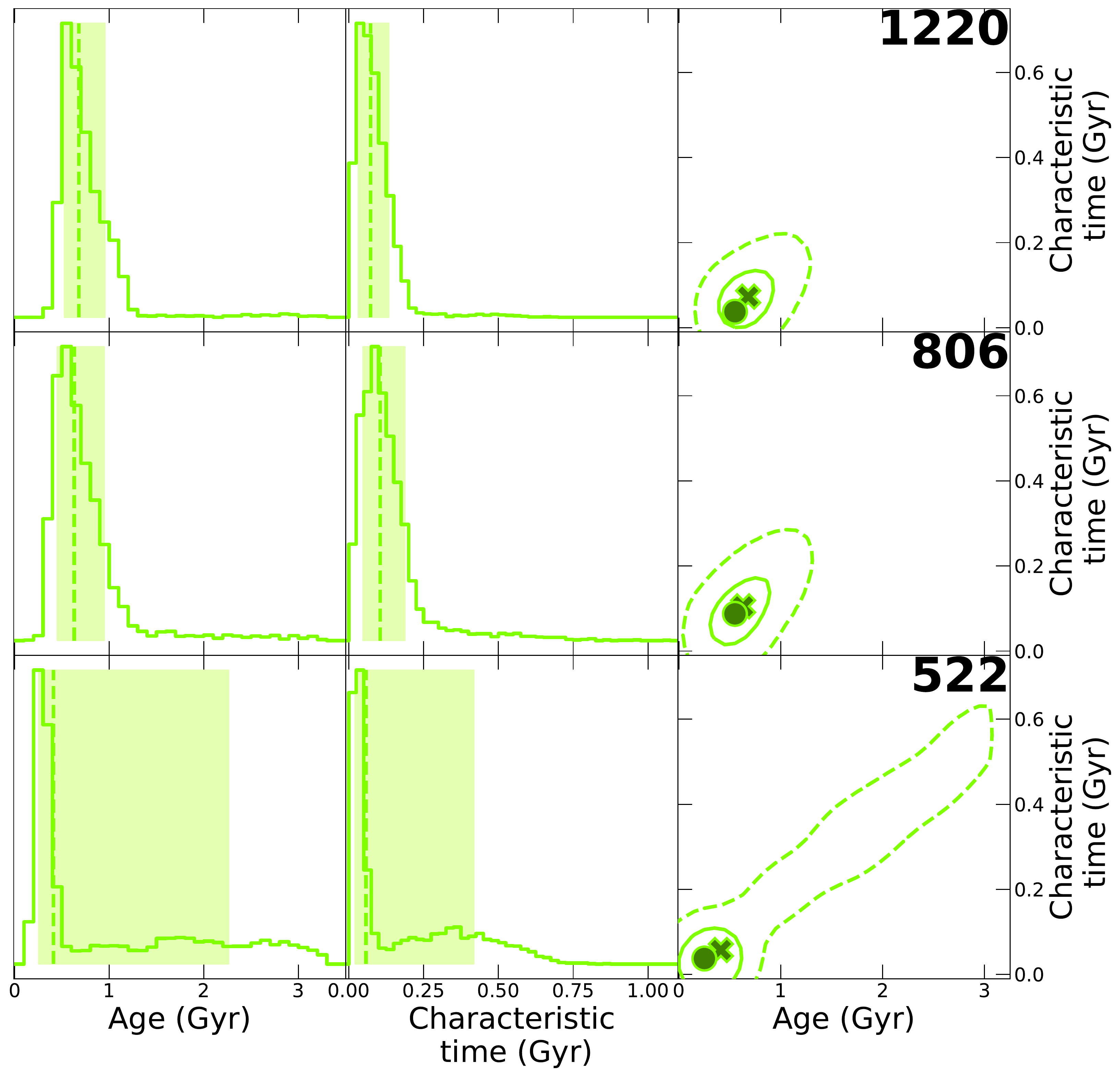}

\caption{Posterior distributions of the ages and characteristic times of the young members, along with their degeneracies. Lines and contours definition are given in Figure \ref{fig_very_old_members}.}
\label{fig_young_members}
\end{figure}

\begin{figure}%[h!]
\centering
\includegraphics[width=8cm]{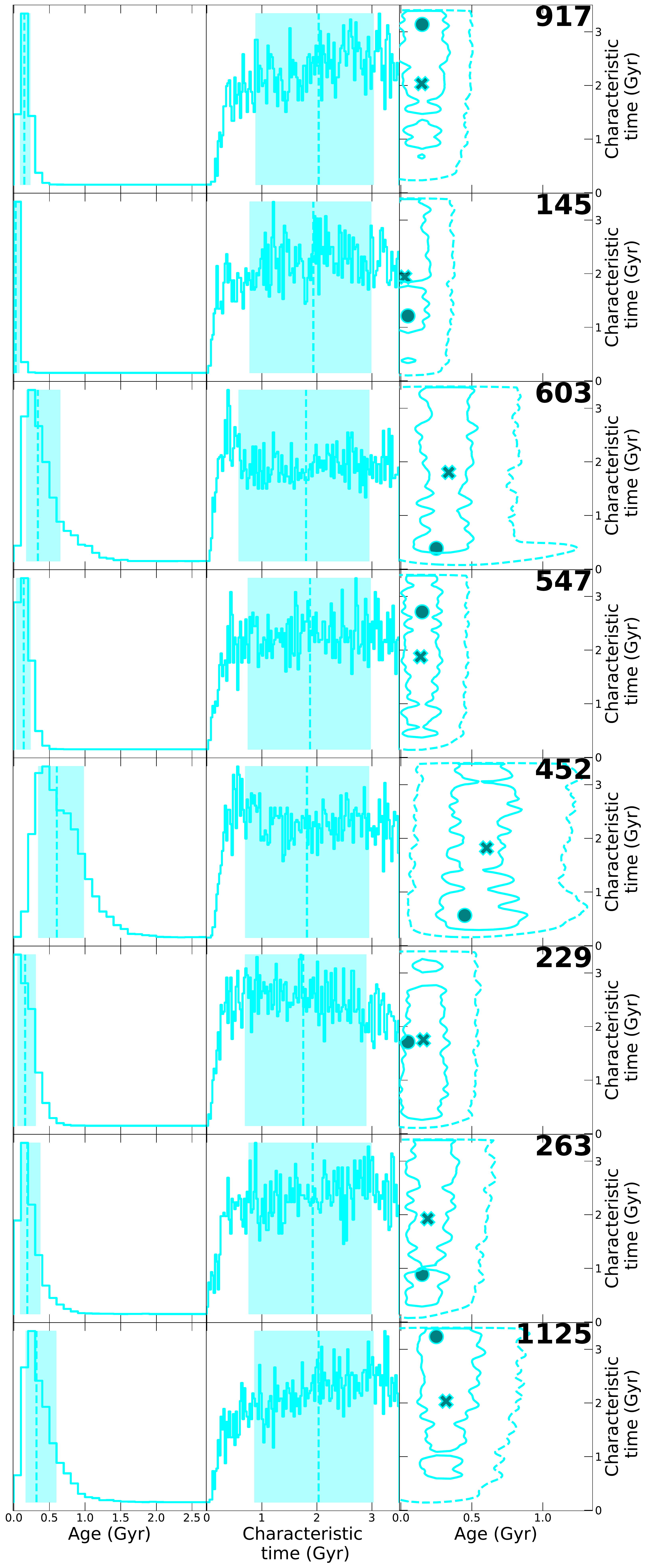}

\caption{Posterior distributions of the ages and characteristic times of the star-forming members, along with their degeneracies. Lines and contours definition are given in Figure \ref{fig_very_old_members}.}
\label{fig_star-forming_members}
\end{figure}

Throughout this paper, we will refer to the members of the very old, old and young age categories as the evolved members. The age categories of each member are presented in Table \ref{table_members} along with their coordinates, redshift, \textit{F140W} magnitude and quality category. For convenience, we adopt \cite{willis_spectroscopic_2020} IDs. Figures \ref{fig_very_old_members} to \ref{fig_star-forming_members} show the posterior distributions of ages and characteristic times, as well as their degeneracies. Four examples of a complete set of parameters distributions and degeneracies are given in the Appendix \ref{sec_corner_plots}.

The oldest members, presented in Figure \ref{fig_very_old_members}, are characterised by relatively broad age posterior distributions, with the $1\sigma$ confidence region usually covering the interval between 1.5 and 3 Gyrs. Characteristic time posterior distribution are consistent with $\tau \lesssim$ 0.5 Gyr, although they tend to be broader than for other evolved members, due to the degeneracies with the age distributions. In the top panel of Figure \ref{fig_XLSSC122_members}, these members are concentrated within the cluster core: the average projected distance to the BCG is 147 kpc (for the old, young and star-forming members the average distances are respectively 380 kpc, 473 kpc and 347 kpc). In contrast to this, old members, shown in Figure \ref{fig_old_members} present various intermediate posterior distributions between the broad age posteriors of the oldest members and the peaked ones of the young members. The $\tau$ distributions suggest short characteristic times, usually approximately equal or smaller than 0.3 Gyr. Both the young and star-forming members (Figures \ref{fig_young_members} and \ref{fig_star-forming_members}) are younger than 0.75 Gyr. Young and star-forming members are differentiated by the shape of their $\tau$ posterior distributions: shorter than 0.2 Gyr and degenerate with the age distribution for the young members; very poorly constrained and independent from the age distribution (or any other fitted parameters) for the star-forming members. Their location on the CMD are diagram are also different: the star-forming members populate the blue cloud while the young members reside in the red sequence, with the exception of ID 522.

\begin{table*}
\caption{Member ID, position, quality of the fit and age categories. The medians of the age, characteristic time and stellar mass distributions are also presented.}%\textcolor{cyan}{The medians of the posteriors of the ages of the oldest stars and the medians of characteristic time distributions are also presented.}}
\label{table_members}
\centering
%\setlength{\tabcolsep}{2pt}
%%\tablenum{1}
\begin{tabular}{c c c c c c c c c c}
\hline
ID$^a$ & RA$^b$ & Dec $^b$ & Redshift$^a$ & \textit{F140W} magnitude & Quality & Age class & Median age & Median $\tau$ & Median mass$^c$\\
 & (deg) & (deg) &  &  &  &  & (Gyr) & (Gyr) & (log$_{10}$(M/M$_\odot$)\\
 \hline
526 & 34.4342 & -3.7588 & 1.98 & $\mathrm{22.09\pm 0.02}$ & silver & very old & 2.24 & 0.28 & 11.20\\
451 & 34.4223 & -3.7635 & 1.98 & $\mathrm{22.31\pm 0.02}$ & gold & old & 1.29 & 0.24 & 10.92\\
657 & 34.4341 & -3.7577 & 1.98 & $\mathrm{22.53\pm 0.02}$ & silver & very old & 2.33 & 0.27 & 11.02\\
295 & 34.4353 & -3.7680 & 1.99 & $\mathrm{22.67\pm 0.02}$ & gold & very old & 2.49 & 0.34 & 11.00\\
1032 & 34.4325 & -3.7499 & 1.98 & $\mathrm{22.69\pm 0.02}$ & gold & old & 1.64 & 0.09 & 10.74\\
240 & 34.4224 & -3.7700 & 1.98 & $\mathrm{22.73\pm 0.03}$ & gold & old & 0.96 & 0.06 & 10.47\\
1064 & 34.4359 & -3.7495 & 1.99 & $\mathrm{22.82\pm 0.03}$ & bronze & - & -  & - & - \\
917 & 34.4356 & -3.7531 & 1.96 & $\mathrm{22.83\pm 0.03}$ & gold & star-forming & 0.15 & 2.03 & 10.01\\
298 & 34.4472 & -3.7680 & 1.99 & $\mathrm{22.90\pm 0.03}$ & gold & old & 0.93 & 0.07 & 10.60\\
1050 & 34.4369 & -3.7502 & 1.98 & $\mathrm{22.99\pm 0.03}$ & gold & old & 1.68 & 0.12 & 10.73\\
606 & 34.4385 & -3.7607 & 1.97 & $\mathrm{23.05\pm 0.03}$ & silver & very old & 2.31 & 0.47 & 10.76\\
236 & 34.4516 & -3.7703 & 1.98 & $\mathrm{23.30\pm 0.07}$ & bronze & - & -  & - & - \\
642 & 34.4338 & -3.7588 & 2.04 & $\mathrm{23.31\pm 0.03}$ & bronze & - & -  & - & - \\
145 & 34.4448 & -3.7729 & 1.98 & $\mathrm{23.32\pm 0.03}$ & gold & star-forming & 0.03 & 1.94 & 9.74\\
372 & 34.4441 & -3.7657 & 1.96 & $\mathrm{23.43\pm 0.04}$ & bronze & - & -  & - & - \\
402 & 34.4464 & -3.7653 & 1.97 & $\mathrm{23.46\pm 0.04}$ & bronze & - & -  & - & - \\
734 & 34.4250 & -3.7580 & 2.00 & $\mathrm{23.48\pm 0.04}$ & gold & very old & 1.86 & 0.12 & 10.60\\
845 & 34.4347 & -3.7549 & 1.98 & $\mathrm{23.51\pm 0.04}$ & gold & very old & 2.47 & 0.28 & 10.43\\
1220 & 34.4434 & -3.7450 & 1.98 & $\mathrm{23.53\pm 0.04}$ & silver & young & 0.68 & 0.07 & 10.10\\
493 & 34.4330 & -3.7632 & 1.96 & $\mathrm{23.67\pm 0.04}$ & silver & very old & 2.21 & 0.17 & 10.52\\
649 & 34.4340 & -3.7593 & 2.00 & $\mathrm{23.74\pm 0.04}$ & bronze & - & -  & - & - \\
603 & 34.4394 & -3.7603 & 1.98 & $\mathrm{23.77\pm 0.04}$ & gold & star-forming & 0.34 & 1.80 & 9.72\\
345 & 34.4419 & -3.7667 & 1.99 & $\mathrm{23.77\pm 0.04}$ & silver & old & 1.62 & 0.33 & 10.12\\
1141 & 34.4336 & -3.7478 & 1.96 & $\mathrm{23.88\pm 0.04}$ & bronze & - & -  & - & - \\
730 & 34.4398 & -3.7583 & 1.99 & $\mathrm{23.98\pm 0.05}$ & gold & very old & 2.30 & 0.44 & 10.52\\
547 & 34.4353 & -3.7625 & 1.96 & $\mathrm{24.06\pm 0.05}$ & gold & star-forming & 0.14 & 1.88 & 9.24\\
452 & 34.4190 & -3.7639 & 1.97 & $\mathrm{24.07\pm 0.06}$ & silver & star-forming & 0.60 & 1.82 & 9.63\\
229 & 34.4405 & -3.7704 & 1.98 & $\mathrm{24.18\pm 0.05}$ & silver & star-forming & 0.16 & 1.75 & 9.46\\
726 & 34.4306 & -3.7576 & 1.97 & $\mathrm{24.18\pm 0.05}$ & gold & dusty$^d$ & 2.22 & 1.55 & 10.60\\
806 & 34.4477 & -3.7561 & 1.98 & $\mathrm{24.19\pm 0.05}$ & gold & young & 0.63 & 0.11 & 9.71\\
1253 & 34.4463 & -3.7436 & 2.02 & $\mathrm{24.27\pm 0.08}$ & bronze & - & -  & - & - \\
466 & 34.4187 & -3.7637 & 1.98 & $\mathrm{24.27\pm 0.07}$ & bronze & - & -  & - & - \\
428 & 34.4466 & -3.7645 & 1.98 & $\mathrm{24.35\pm 0.06}$ & bronze & - & -  & - & - \\
329 & 34.4276 & -3.7674 & 1.97 & $\mathrm{24.40\pm 0.06}$ & bronze & - & -  & - & - \\
263 & 34.4211 & -3.7692 & 1.98 & $\mathrm{24.40\pm 0.07}$ & silver & star-forming & 0.19 & 1.93 & 9.50\\
522 & 34.4190 & -3.7628 & 1.96 & $\mathrm{24.49\pm 0.06}$ & silver & young & 0.41 & 0.06 & 9.53\\
1125 & 34.4387 & -3.7482 & 2.00 & $\mathrm{24.49\pm 0.06}$ & silver & star-forming & 0.32 & 2.03 & 9.63\\
\hline
\multicolumn{10}{l}{$^a$ IDs and spectroscopic redshifts are from \citet{willis_spectroscopic_2020}.}\\        
\multicolumn{10}{l}{$^b$ Positions are based on the \textit{F140W} band astrometry.}\\
%\multicolumn{9}{l}{$^c$ The only galaxy that doesn't fit into a category. See Appendix \ref{ssec_726} for more details.}\\
\multicolumn{10}{l}{$^c$ This is the current stellar mass, while the fitted free parameter is the total stellar mass formed during the star formation history.}\\
\multicolumn{10}{l}{$^d$ The only galaxy that doesn't fit into a category. See Appendix \ref{ssec_726} for more details.}\\
\end{tabular}
\end{table*}
%[OIII] doublet at 4959 and 5007angstrom: i.e. 14778 and 14921 angstrom; Hβ at 14487 A (4861 A), Hγ at 12933 (4340).

We note however that while the exponentially decreasing modelling provides an easy-to-interpret overview of the star formation history in XLSCC 122 members, detailed star formations histories are likely to be more complex and varied \citep[e.g.][]{sparre_star_2015,leja_how_2019,tacchella_fast_2022}. 
For example, mergers might generate more complex star formation histories. Massive clusters members are often the product of dry or wet mergers \citep{brodwin_era_2013,cattaneo_star-forming_2013,cooke_formation_2015,wagner_star_2015} which might bring together galaxies with different star formation histories, or, in the case of wet mergers, trigger a temporary increase of the star-forming rate.

Another element to consider is the impact of the assumed prior on to the recovered star formation histories. For example, \citep{carnall_how_2019} shows that simple $\tau$-models, such as the ones employed here, struggle to recover constant or rising star formation rates, which might explain why the characteristic times of our star-forming members are so poorly constrained. To overcome the limitations of simple $\tau$-models, several ideas have been proposed: delayed $\tau$-models \citep[e.g.][]{pacifici_rise_2013,simha_parametrising_2014}, log-normal models \citep[e.g.][]{diemer_log-normal_2017}, double power laws \citep[e.g.][]{carnall_inferring_2018} or non-parametric modelling \citep[e.g.][]{iyer_nonparametric_2019,leja_how_2019} constitute a non-exhaustive list.

\subsection{Testing the dependence upon the assumed star formation model}\label{ssec_delayed_tau}

\begin{figure}%[h!]
\centering
\includegraphics[width=8cm]{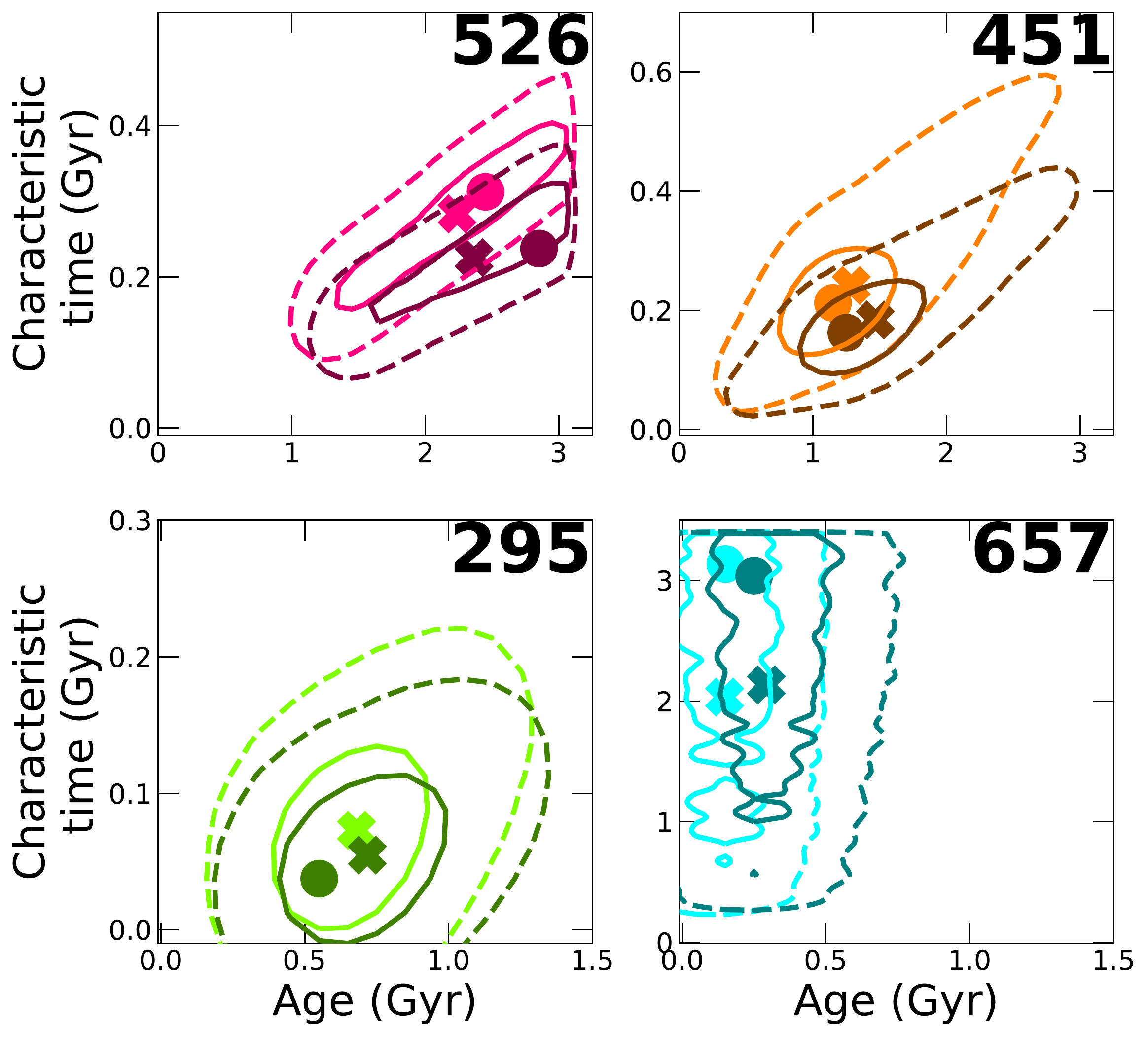}

\caption{Comparison between the age-characteristic time degeneracies of the simple and delayed $\tau$-models, for representative members of the four age categories. The simple $\tau$-model degeneracies follow the colour scheme introduced in Figure \ref{fig_XLSSC122_members} (pink for very old, orange for old, green for young and cyan for star-forming) with the delayed $\tau$-model degeneracies overlaid in darker colours. The 1 and 2$\sigma$ degeneracies are represented by full and dashed contours respectively.}
\label{fig_comparison_tau_delayed}
\end{figure}

In order to assess the dependence of our results on the assumed star formation history, we tested the effect of a change of the model. \textsc{Bagpipes} offers six basic parametric star formation histories, ranging from a total of four to six free parameters (assuming that the stellar mass, the metal and dust contents are allowed to vary). More complex models can be build by combining several parametric models together \citep[for more detail, see Sections 3.1.2 and 4.2 of][]{carnall_inferring_2018}. However, given the quality of our data, a six-parameter model such as a double power law \citep{behroozi_average_2013,carnall_inferring_2018} tends to generate poorly constrained results, especially for the less luminous members of the red sequence. 

At the exception of the constant, rectangularly-shaped star formation history and the currently used $\tau$-model, all of the five parameters models available rely on the assumption that the star formation declines more slowly than it rises. Among those, the delayed $\tau$-model has the form:

\begin{equation}\label{eq_delayed_tau_time}
\mathrm{SFR(t)}\propto\left\{
\begin{array}{ll}
(t-t_0) e^{-(t-t_0)/\tau} ~&\mathrm{if} ~t\geq t_0\\
0&\mathrm{else}
\end{array}
\right.
\end{equation}

\noindent where $t_0$ is the formation time of the oldest stars and $\tau$ is the characteristic time, or

\begin{equation}\label{eq_delayed_tau_age}
\mathrm{SFR(a)}\propto \left\{
\begin{array}{ll} 
(a_0-a) e^{-(a_0-a)/\tau}~&\mathrm{if} ~a \leq a_0\\
0&\mathrm{else}
\end{array}
\right.
\end{equation}

\noindent when expressed in term of ages. In this latter form, $a_0$ is the age of the oldest stars. 

The similarities between the parameters of the simple and delayed $\tau$-model allow us to compare the results of the two models directly. The results of such a comparison are presented in Figure \ref{fig_comparison_tau_delayed} for four representative galaxies. For the evolved members, the distributions of the age of oldest stars remain similar between the models but the characteristic times of the delayed model are slightly shifted toward lower values. For the star-forming members, we observe a shift toward higher ages.

This behaviour can be understood if one notes that the SED photometric data essentially constrains the duration and last epoch of major star formation in any model fit. The passive nature of the evolved members places a firm constraint on the age at which star formation must end in these systems. The delayed $\tau$-model differs from the simple $\tau$-model in that the star formation rate takes an amount of time equivalent to the value of $\tau$ to rise from zero star formation to a maximum value, from which it then declines. As the SED photometric data constrains the duration of star formation in the fitted model, the effect of adding this early time behaviour to the model (effectively an additional $\tau$ of rising star formation) is compensated for in the fitting procedure by reducing the overall $\tau$ value of the model. In this sense the slightly different results generated by the two star formation models can be understood. Each model generates fits of equivalent statistical merit and, with no direct evidence available to constrain the form of the SFR(t), we adopt the simple $\tau$-model as our baseline model yet note the minor numerical differences between the two models that propagate through our later analyses as they arise in the paper.

\subsection{Mock photometry fits for two population SEDs}\label{ssec_mock_photometry}

To explore the effects upon the SED fitting process of the presence of more than one stellar population in a galaxy, we created mock observations characterised by two episodes of star formation with exponentially declining star-forming rates. We then fitted them with a simple $\tau$-model, as described in Section \ref{ssec_fitting}. The first episode starts 2 Gyrs ago and possesses a characteristic time of 300 Myr; the other begins 150 Myr ago and has $\tau=2$ Gyrs. The contribution of the second episode varies from 0.2 to 40 per cent of the total stellar mass formed, which is fixed to $10^{11}~\mathrm{M_\odot}$. Metallicity is set to $\mathrm{Z_\odot}$ and dust extinction to 0.8 magnitude. Photometric errors are estimated by calculating the average photometric errors of the very old and star-forming members. Those averages are multiplied by the fraction of \lq old\rq~and \lq young\rq~components and then summed. Figures are presented in Appendix \ref{sec_mock_photometry_plot}. We found that when the younger component accounts for 4 per cent or more of the total stellar mass, the recovered $\tau$ posterior is poorly constrained, like in the star-forming members. Additionally, when the younger component represents $\gtrsim$ 20 per cent of the stellar mass, the recovered median age starts to be consistent with our criterion for star-forming members.

This allow us to place further constraints on the stellar population of each age category. Very old members contain less than 4 per cent of young stars: otherwise the confidence interval of their characteristic time posterior would be one Gyr broad or more. For the same reason, old and young members are unlikely to host significant populations of newly formed stars. In contrast, star-forming member members might possess up to 80 per cent of older stars.

\section{Discussion}\label{sec_discussion}

\subsection{XLSSC 122 star formation history and its implications}\label{ssec_discussion_sfh}

\begin{figure}
\centering
\includegraphics[width=8cm]{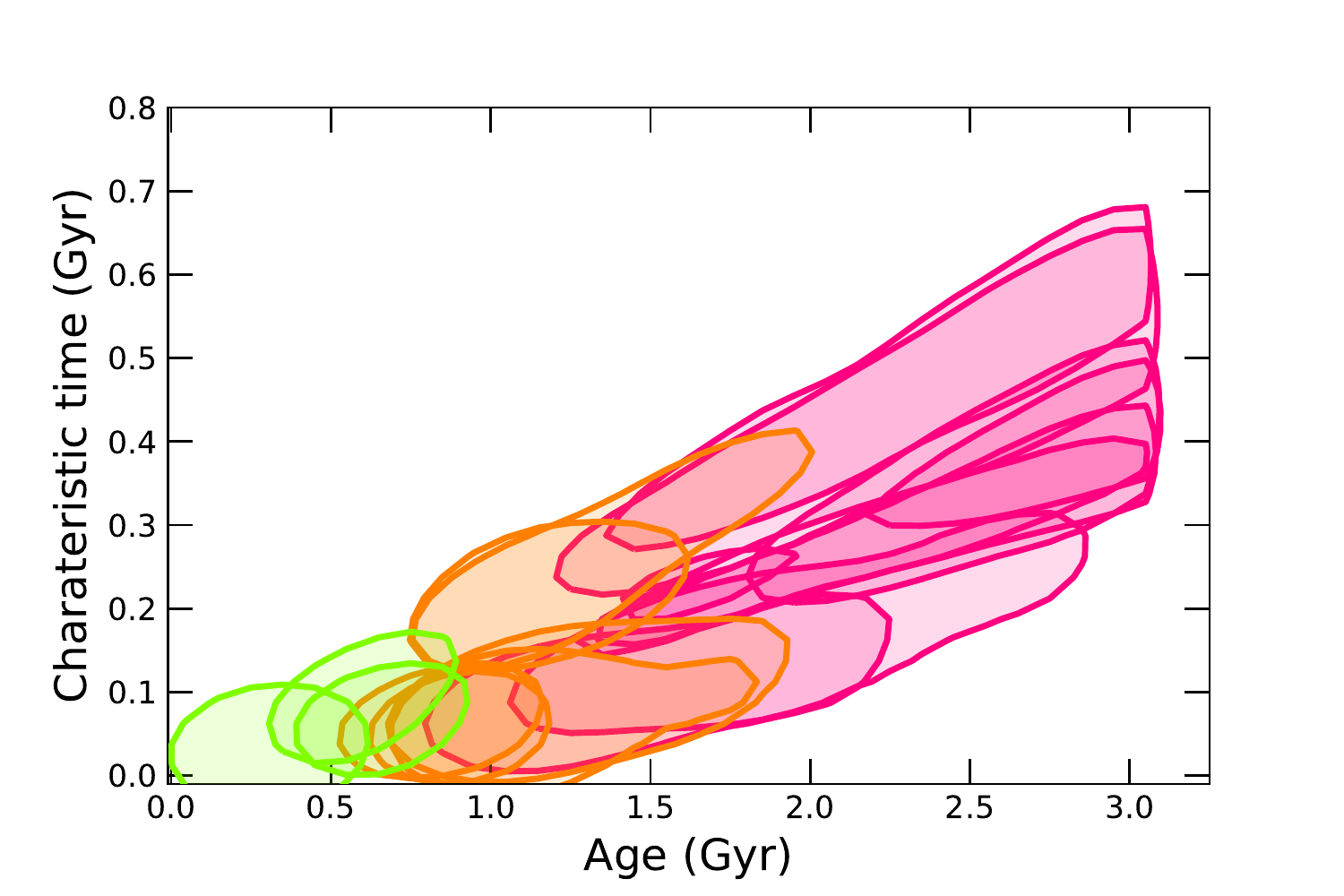}

\caption{The 1$\sigma$ degeneracies between age and characteristic time, for each evolved members. Same colour scheme as before.}
\label{fig_evolution_age_tau}
\end{figure}

The bottom panel of Figure \ref{fig_XLSSC122_members} shows that the colour bimodality observed in XLSSC 122 corresponds to different star formation histories for the red sequence and the blue cloud. The former displays ages varying from approximately 0.5 to 3 Gyrs and short characteristic time-scales while the latter is more uniform, with ages younger than $\sim$ 0.6 Gyr and, at the exception of ID 522, poorly constrained characteristic times.

This relative homogeneity might be explained by different factors. Several authors \citep[e.g.][]{li_how_2007,carnall_how_2019} found that parametric stellar population models tend to be biased toward younger stellar ages, especially if more than one population are present. Our own tests, presented in Section \ref{ssec_mock_photometry}, show that above a certain percentage of young stars, it becomes very difficult to determine whether an older component is present of not. Thus, we cannot determine if our star-forming members experienced a prolonged star formation activity.

For the red-sequence members, the characteristic time distributions are short and relatively well-constrained. This observation argues against the occurrence of more than one major epoch of star formation per galaxy, despite the variety of ages observed. We note however that the oldest galaxies age posteriors are broad. This might either indicates the presence of two or several evolved stellar populations close in ages, perhaps brought together by mergers \citep[e.g.][]{cooke_formation_2015}, or arises from the difficulty of constraining the ages of evolved stellar populations, especially when the available observations place only limited constrains on the metallicity and dust content \citep[e.g.][see also the examples of full parametric distributions in Appendix \ref{sec_corner_plots}]{conroy_modeling_2013,carnall_timing_2020}. The inclusion of near-infrared and/or mid-infrared data with a good resolution \citep{conroy_modeling_2013,carnall_vandels_2019} would help to further constrain the fits; photometric or spectroscopic observations from the newly launched \textit{James Webb Space Telescope} would be ideal in that regard. More realistically, medium-bandwidth photometry in the \textit{H} and \textit{J} filter \citep[e.g.][]{straatman_fourstar_2016} would increase our coverage of the SED close to the 4000 \AA~break.

Figure \ref{fig_evolution_age_tau} shows the 1$\sigma$ degeneracy between age and characteristic time, for the evolved members. There is a distinct trend: the younger the member the shorter the characteristic time. This observation suggests an increase of the quenching efficiency with time, possibly due to the cluster mass increase over time. There is a mixed set of precedents: some authors find an increase of quenching efficiency with time in high-redshift groups and clusters \citep[e.g.][]{nantais_stellar_2016,nantais_evidence_2017,kawinwanichakij_effect_2017} while others found a decrease \citep[e.g.][]{balogh_evidence_2016,foltz_evolution_2018} or a constant efficiency \citep[e.g.][]{lemaux_persistence_2019}. These discrepant measurements might point toward a dependence of the quenching efficiency on the halo mass, i.e. larger structures being more effective in quenching galaxies than groups \citep[e.g.][]{peng_mass_2010,peng_mass_2012,balogh_evidence_2016,foltz_evolution_2018}, although this is again controversial \citep[e.g.][]{fossati_galaxy_2017}.

\subsection{Link with the cluster assembly history}\label{ssec_assembly}

\begin{figure}
\centering
\includegraphics[width=8cm]{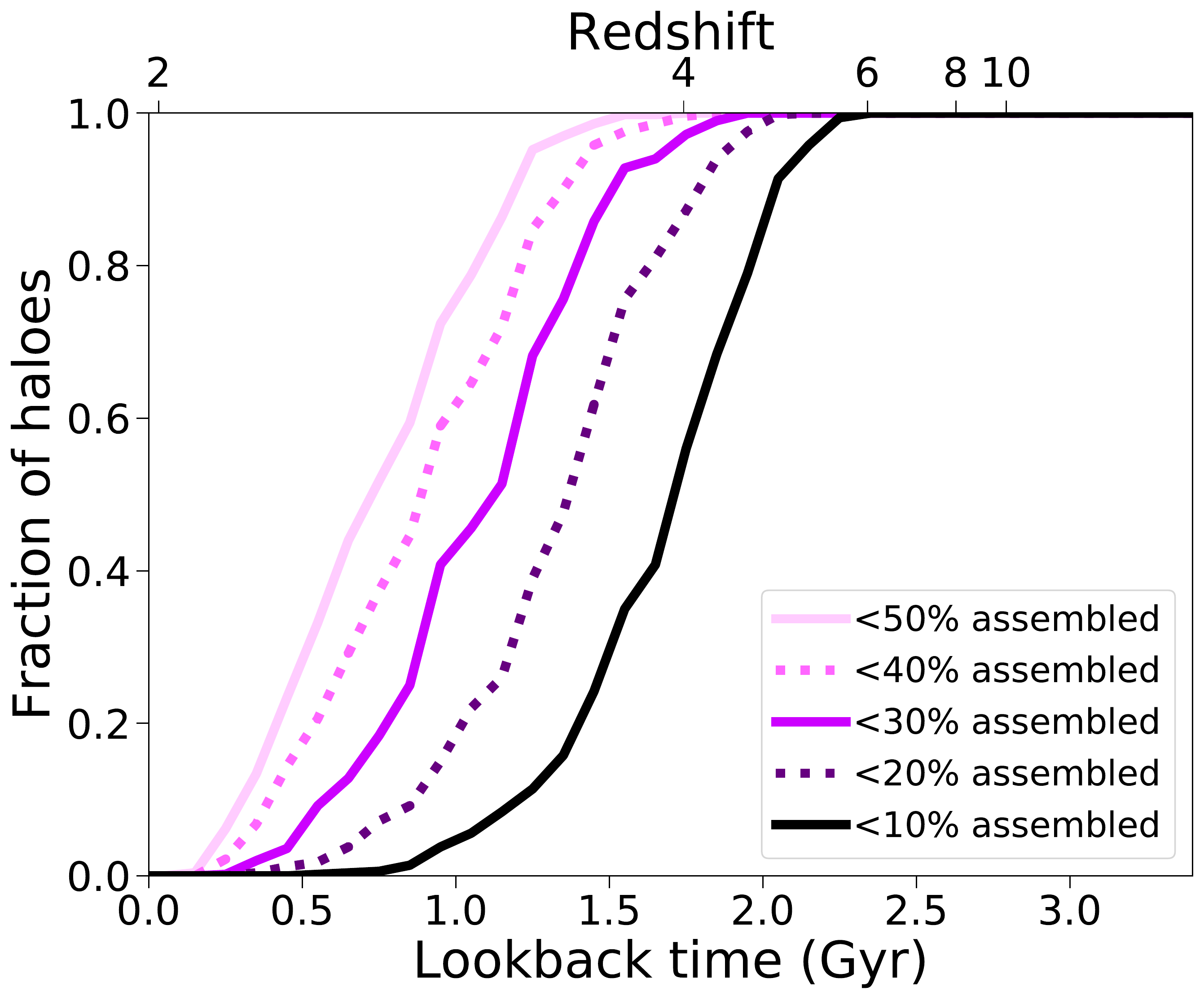}
\caption{Evolution with the time of the fraction of the XLSSC 122-like simulated haloes that are less than 10\% assembled, 10 to 20\% assembled, etc. Simulated haloes assemblies are colour-coded, from black (less than 10\% assembled) to pale lilac (more than 50\%) assembled}
\label{fig_DM_haloes}
\end{figure}

To investigate XLSSC 122 assembly history, we select 500 dark matter haloes in the MultiDark Planck 2 simulation \citep{klypin_multidark_2016}, with $\mathrm{M_{500}}$ masses between $4.3\times 10^{13} ~M_\odot$ and $8.3\times 10^{13} ~M_\odot$ at $z=1.98$. This mass interval is centred on our best estimate of the XLSSC 122 mass, $6.3\times 10^{13}~M_\odot$, and is likely to encompass its true mass as it is slightly larger than the 1$\sigma$ confidence interval~$\pm 1.5 \times 10^{13}~M_\odot$. We then determine the percentage of its $z=1.98$ mass accreted by each halo at any earlier epochs. We organize the results by intervals of 10 per cent, from less than 10 per cent to more than 90 per cent assembled. Figure \ref{fig_DM_haloes} shows the evolution with time of the fraction of haloes that are less than 50 per cent assembled.

Before $z=6$ (2.34 Gyrs before $z=1.98$), none of the dark matter haloes had accreted more than 10 per cent of its z=1.98 mass, in contrast to the wide range of accreted fractions possible at $z=3.5$ (1.48 Gyrs before $z=1.98$). This variety of accreted fractions at later times is representative of the variety of the assembly histories: some haloes will reach their final mass via multiples minor merging events, while others will experience a few major mergers, sometimes accreting more than 30 per cent of their $z=1.98$ masses in a single time step.

\begin{figure*}
\centering
\includegraphics[width=16.5cm]{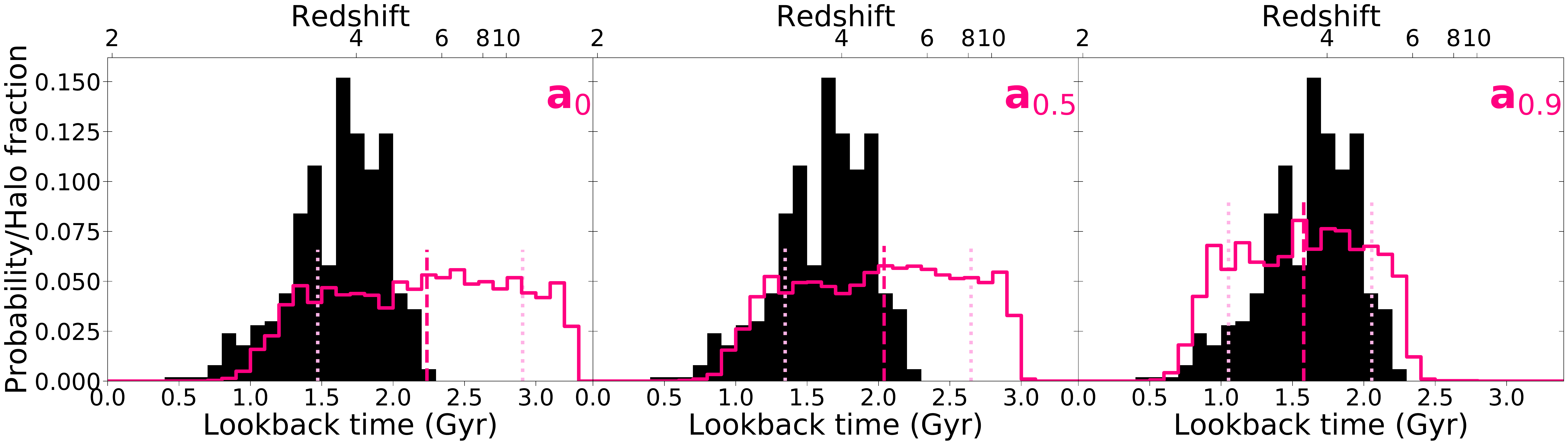}
\caption{\textit{Left}: Comparison between the time at which each simulated halo becomes 10 per cent assembled or more (black histogram) and the posterior of the age of the oldest stars in the BCG. The posterior is traced in pink, and its median is indicated by a pink dashed line. The dotted light pink lines show the edges of its 1$\sigma$ confidence interval. \textit{Centre}: Comparison between the time at which each halo becomes 10 per cent assembled and the distribution corresponding to the time-scale at which the BCG formed 50 per cent of its stellar mass (i.e. $a_{0.5}$). \textit{Right}: Comparison between the time at which each halo becomes 10 per cent assembled and the $a_{0.9}$ distribution.}
\label{fig_assembly_526}
\end{figure*}

The left panel of Figure \ref{fig_assembly_526} presents a comparison between the age posterior distribution of the BCG ($a_0$) with the time at which each simulated halo becomes at least 10 per cent assembled. The 10 per cent assembly times for the simulated haloes correspond roughly to the younger half of the posterior distribution, showing that the onset of star formation occurred long before virialization.

We then quantitatively compare the fitted star formation histories of the member galaxies with the mass assembly histories of the simulated haloes by weighting them by the age posterior. For example, the weighted fraction of haloes that have accreted less than 10 per cent of their $z=1.98$ masses is given by:

\begin{equation}\label{eq_composed_prob}
f_{form,<10\%}=\sum_{t} P_{age}(t) \times f_{<10\%}(t),
\end{equation}

\noindent
where $P_{age}(t)$, the probability that star formation started at time t, is given by the value at t of the normalized age posterior of the galaxy. The term $f_{<10\%}(t)$ is the fraction of simulated haloes that have accreted less than 10 per cent of their $z=1.98$ masses at t. We can similarly compute $f_{form,10-20\%}$, the weighted fraction of haloes that have accreted between 10 and 20 per cent of their mass, and so on. Table \ref{table_assembly_formation_time} shows the weighted fraction of haloes that have accreted less than 10, 20, 30 or 40 per cent of their masses at the time of the onset of the star formation in the very old cluster members. The bottom line shows that, on average, 74 per cent of the haloes had accreted less than 10 per cent of their mass when the very old members began to form their stars. 

The $\tau$-model used to determine the star formation histories does not allow us to directly determine a quenching time. Instead, we use the time-scales at which cluster members formed 50 and 90 per cent of their stellar masses as proxies to place this event in the cluster accretion history. To calculate $t_{X}$, the time when a fraction X of the stellar mass was formed, one must solve the following equation:

\begin{equation}\label{eq_integral_t_X}
\int_{t_0}^{t_{X}} \mathrm{SFR} dt=X\int_{t_0}^{t_{obs}} \mathrm{SFR} dt,
\end{equation}

\noindent
where $t_0$ is the formation time of the oldest stars and $t_{obs}$ the epoch at which we observe the galaxy. Thus, the age posterior is given by $a_0=t_{obs}-t_0$. Due to the simplicity of the $\tau$-model this equation has an analytical solution:

\begin{equation}\label{eq_t_X}
t_{X}-t_0=-\tau ln[Xe^{-(t_{obs}-t_0)/\tau}-X+1].
\end{equation}

%\noindent

We then define $a_{X}=t_{obs}-t_X$ the time since a fraction X of the star were formed, and reorganise the equation to get:

\begin{equation}\label{eq_a_X}
a_X=\tau ln[Xe^{-(a_0)/\tau}-X+1]+a_0.
\end{equation}

The central and right panels of Figure \ref{fig_assembly_526} compare the times at which each simulated halo becomes at least 10 per cent assembled to respectively the $a_{0.5}$ and $a_{0.9}$ \lq posteriors\rq~for the BCG. The BCG acquires 90 per cent of its stellar mass on approximately the same time-scale for the simulated haloes to become 10 per cent assembled.

By replacing $P_{age}(t)$ in Equation \ref{eq_composed_prob} by probabilities drawn from the $a_{0.5}$ and $a_{0.9}$ posteriors, we compute the fraction of haloes less than 10 per cent assembled when the first galaxies had formed 50 or 90 per cent of their stellar masses. Results are given in Tables \ref{table_assembly_50_time} and \ref{table_assembly_90_time}. We thus determine that 67 per cent of the haloes had accreted 10 per cent of their masses when the oldest members had 50 per cent of their own stellar masses formed. Although there is more variability than for the onset of star formation, we also compute that when 90 per cent of the first galaxies stellar masses were in place, 42 per cent of the simulated haloes had accreted less than 10 per cent of their masses. This percentage rises to 74 per cent for the haloes that are less than 30 per cent assembled.

Changing the assumed star formation history to a delayed $\tau$-model tends to decrease the values of the $a_{0.5}$ and $a_{0.9}$ distributions. However the impact is limited: with a delayed $\tau$-model, an average of 69 per cent of the simulated haloes had accreted 30 per cent of their masses when the oldest members had 90 per cent of their own stellar masses formed. Appendix \ref{sec_delayed_assembly_history} presents further assembly predictions for this model, and a comparison of the $a_0$, $a_{0.5}$ and $a_{0.9}$ posteriors obtained for the BCG with the two star formation histories.

Our results suggest that the oldest members in XLSSC 122 formed and were quenched in the protocluster stage, in agreement with previous attempts to link star formation history and overdensity evolution \citep[e.g.][]{poggianti_evolution_2006,muldrew_galaxy_2018,werner_satellite_2022}. Our conclusions are also supported by recent observations of $z\sim 3$ protoclusters already harbouring quenched galaxies \citep[e.g.][]{shi_census_2019,shi_accelerated_2021,kalita_ancient_2021,kubo_massive_2021}, which might have formed around $z\sim 4$ \citep[e.g.][]{long_emergence_2020}.

\begin{table*}%[!ht]
\caption{Cumulative fractions of XLSSC 122 haloes that are $<$10\% assembled, $<$20\% assembled, etc., integrated over the age distributions of the oldest members.}
\label{table_assembly_formation_time}
\centering
%\setlength{\tabcolsep}{2pt}
%%\tablenum{1}
\begin{tabular}{c c c c c}
\hline
ID & $<$10\% & $<$20\% & $<$30\% & $<$40\%\\
 %\\
 \hline
526 & 0.75 & 0.87 & 0.94 & 0.97\\
657 & 0.74 & 0.85 & 0.91 & 0.95\\
295 & 0.78 & 0.88 & 0.93 & 0.96\\
606 & 0.75 & 0.86 & 0.92 & 0.95\\
734 & 0.60 & 0.74 & 0.84 & 0.91\\
845 & 0.81 & 0.88 & 0.92 & 0.95\\
493 & 0.72 & 0.84 & 0.91 & 0.95\\
730 & 0.75 & 0.85 & 0.91 & 0.94\\
average & 0.74 & 0.85 & 0.91 & 0.95\\
\hline
\end{tabular}
\end{table*}

\begin{table*}%[!h]
\caption{Cumulative fraction of XLSSC 122-like haloes that are partially assembled, integrated over the time distributions corresponding to the formation of 50\% of the stellar masses of the oldest members.}
\label{table_assembly_50_time}
\centering
%\setlength{\tabcolsep}{2pt}
%%\tablenum{1}
\begin{tabular}{c c c c c c}
\hline
ID & $<$10\% & $<$20\% & $<$30\% & $<$40\% & $<$50\%\\
 %\\
 \hline
526 & 0.68 & 0.83 & 0.91 & 0.95 & 0.98\\
657 & 0.69 & 0.81 & 0.89 & 0.93 & 0.96\\
295 & 0.73 & 0.84 & 0.90 & 0.94 & 0.96\\
606 & 0.64 & 0.78 & 0.87 & 0.92 & 0.95\\
734 & 0.56 & 0.71 & 0.82 & 0.89 & 0.93\\
845 & 0.77 & 0.86 & 0.91 & 0.94 & 0.96\\
493 & 0.68 & 0.81 & 0.89 & 0.94 & 0.96\\
730 & 0.64 & 0.77 & 0.86 & 0.91 & 0.94\\
average & 0.67 & 0.80 & 0.88 & 0.93 & 0.96\\
\hline
\end{tabular}
\end{table*}

\begin{table*}%[!h]
\caption{Cumulative fraction of XLSSC 122-like haloes that are partially assembled, integrated over the time distributions corresponding to the formation of 90\% of the stellar masses of the oldest members.}
\label{table_assembly_90_time}
\centering
%\setlength{\tabcolsep}{2pt}
%%\tablenum{1}
\begin{tabular}{c c c c c c c c}
\hline
ID & $<$10\% & $<$20\% & $<$30\% & $<$40\% & $<$50\% & $<$60\% & $<$70\%\\
 %\\
 \hline
526 & 0.43 & 0.65 & 0.79 & 0.87 & 0.93 & 0.96 & 0.98\\
657 & 0.50 & 0.67 & 0.80 & 0.87 & 0.92 & 0.95 & 0.97\\
295 & 0.50 & 0.68 & 0.80 & 0.87 & 0.91 & 0.94 & 0.97\\
606 & 0.18 & 0.39 & 0.58 & 0.71 & 0.79 & 0.85 & 0.90\\
734 & 0.44 & 0.62 & 0.75 & 0.84 & 0.89 & 0.93 & 0.96\\
845 & 0.55 & 0.74 & 0.84 & 0.90 & 0.94 & 0.96 & 0.97\\
493 & 0.53 & 0.70 & 0.81 & 0.88 & 0.93 & 0.95 & 0.97\\
730 & 0.25 & 0.44 & 0.59 & 0.70 & 0.77 & 0.83 & 0.88\\
average & 0.42 & 0.61 & 0.75 & 0.83 & 0.89 & 0.92 & 0.95\\
\hline
\end{tabular}
\end{table*}

\section{Summary}\label{sec_summary}

We have constructed a photometric catalogue to study the galaxy population of XLSSC 122, a mature galaxy cluster at $z=1.98$. Our catalogue contains aperture-corrected flux densities for 37 spectroscopically confirmed members in 12 bands, covering the near ultraviolet to the mid infrared. The stellar populations of 26 of these members were modelled with a exponentially decreasing star-forming rate. Although we also treated the mass, the metallicity and  the dust extinction as free parameters, we focus our analysis on to the age of the oldest stars and the characteristic time.

We found that the red sequence galaxies exhibit a variety of ages, spanning from 0.5 to 3.26 Gyrs old, which is the age of the Universe at this redshift. Their characteristic times are all short, usually $\lesssim$ 0.3 Gyr, and are thus consistent with a short duration of the star formation. In contrast, all but one of the blue cloud members display poorly constrained $\tau$ and young ages. This suggests these objects are still building a significant amount of their stellar masses, but we are unable to determine whether they possess an older stellar population or not.

Age and $\tau$ appear to be linked among the evolved members, with younger galaxies featuring shorter characteristic times than the older ones. We suggest that quenching efficiency increases with the mass of the cluster, but there is no consensus in the literature on that subject.

Finally, we use 500 dark matter haloes from the MultiDark Planck 2 simulations to reconstruct the mass assembly history of XLSSC 122, allowing us to put constrains on the cluster state when its oldest members formed their first stars, and when they had 50 per cent and 90 per cent of their stellar masses in place. We found that, for the oldest galaxies, 74 per cent of the simulated haloes had accreted less than 10 per cent of their $z=1.98$ masses at the time of the star formation onset. When the oldest galaxies had 50 per cent of their stellar masses formed, 67 per cent of the haloes are less than 10 per cent assembled. Similarly, 90 per cent of the oldest members stellar masses were in place, 75 per cent of the haloes were less than 30 per cent assembled. These probabilities allow us to determine that some of the first galaxies in XLSSC 122 where very likely quenched before the cluster virialization, in agreement with the recent discovery of quenched galaxies in $z\sim 3$ protoclusters.

\section*{Acknowledgements}\label{sec_thanks}

The authors wish to thank Lucio Chiappetti for his careful review of the manuscript. AT is supported by the Natural Sciences and Engineering Research Council of Canada (NSERC) Postgraduate Scholarship-Doctoral Program. JPW acknowledges support from the NSERC Discovery Grant program. DR acknowledges the support of the Natural Sciences and Engineering Research Council of Canada (NSERC), [funding reference number 534263]. REAC and EN acknowledge support from NASA grant number HST-GO-15267.002-A. ACC would like to thank the Leverhulme Trust for their support via the Leverhulme Early Career Fellowship scheme. EN acknowledges support from \textit{Chandra} award GO0-21088X (proposal 21700167). MP acknowledges long-term support from the Centre National d'Etudes Spatiales (CNES). XXL is an international project based around an \textit{XMM-Newton} Very Large Programme surveying two 25 $\mathrm{deg^2}$ extragalactic fields at a depth of $\sim 6\times 10^{-15} ~\mathrm{erg ~cm^{-2} ~s^{-1}}$ in the [0.5-2] keV band for point-like sources. The XXL website is \url{http://irfu.cea.fr/xxl}. %Multi-band information and spectroscopic follow-up of the X-ray sources are obtained through a number of survey programmes, summarised at \url{http://xxlmultiwave.pbworks.com/}. 

%Emil and Becky should confirm, but they may wish to acknowledge support from NASA grant number HST-GO-15267.002-A
% EN acknowledges support from Chandra award GO0-21088X (proposal # 21700167).
%%%%%%%%%%%%%%%%%%%%%%%%%%%%%%%%%%%%%%%%%%%%%%%%%%%
\section*{Data availability}\label{sec_retrieve_data}
%
% 
%\textcolor{red}{The inclusion of a Data Availability Statement is a requirement for articles published in MNRAS. Data Availability Statements provide a standardised format for readers to understand the availability of data underlying the research results described in the article. The statement may refer to original data generated in the course of the study or to third-party data analysed in the article. The statement should describe and provide means of access, where possible, by linking to the data or providing the required accession numbers for the relevant databases or DOIs.}

%The data undelying this article are available at \textcolor{red}{some URL probably \url{https://www.cadc-ccda.hia-iha.nrc-cnrc.gc.ca/en/megapipe/}} (CFHTLS; u, g, r and i bands) and \textcolor{red}{some other URL probably \url{http://archive.eso.org/wdb/wdb/adp/phase3_main/form?collection_name=HAWKI}} (VLT HAWK-I; Y, J and $\mathrm{K_s}$ band). HSC-SSP z band data can be found at \url{https://hsc-release.mtk.nao.ac.jp/doc/} and IRAC data are available at \url{https://irsa.ipac.caltech.edu/data/SPITZER/SWIRE/}. Reduced HST image are available from J. P. Willis (j\href{mailto:jwillis@uvic.ca}{jwillis@uvic.ca}) upon reasonable request.

The data underlying this article are available at \url{https://www.cadc-ccda.hia-iha.nrc-cnrc.gc.ca/en/megapipe/} (CFHTLS; \textit{u}, \textit{g}, \textit{r} and \textit{i} bands) and \url{http://archive.eso.org/wdb/wdb/adp/phase3_main/form?collection_name=HAWKI} (VLT HAWK-I; \textit{Y}, \textit{J} and $K_s$ bands). HSC-SSP \textit{z}-band data can be found at \url{https://hsc-release.mtk.nao.ac.jp/das_cutout/pdr2/} and IRAC data are available at \url{https://irsa.ipac.caltech.edu/data/SPITZER/SWIRE/}. All \textit{HST} data presented in this paper are publicly available in the \textit{Hubble} Legacy Archive (\url{https://hla.stsci.edu/}). The programme number is 15267.

%Reduced HST images are available from J. P. Willis (j\href{mailto:jwillis@uvic.ca}{jwillis@uvic.ca}) upon reasonable request.

%\textcolor{red}{check url HAWK-I data with Jon}

%\url{https://www.cadc-ccda.hia-iha.nrc-cnrc.gc.ca/en/megapipe/cfhtls/cq.html}  https://www.cadc-ccda.hia-iha.nrc-cnrc.gc.ca/en/megapipe/ https://www.cadc-ccda.hia-iha.nrc-cnrc.gc.ca/en/search/?collection=CFHTMEGAPIPE&noexec=true
%http://archive.eso.org/wdb/wdb/adp/phase3_main/form?collection_name=HAWKI
%https://hsc-release.mtk.nao.ac.jp/das_cutout/pdr2/ https://hsc-release.mtk.nao.ac.jp/doc/


\bibliographystyle{mnras}
%\bibliography{references_XLSSC122} % if your bibtex file is called example.bib
\bibliography{Accepted_manuscript_file} % if your bibtex file is called example.bib

%\begin{thebibliography}{99}
%\end{thebibliography}
%
%%%%%%%%%%%%%%%%%%%%%%%%%%%%%%%%%%%%%%%%%%%%%%%%%%%
%
%%%%%%%%%%%%%%%%%% APPENDICES %%%%%%%%%%%%%%%%%%%%%
%
\appendix

\section{Notes on individual galaxies}\label{sec_weirdo}

\subsection{The BCG}\label{ssec_BCG}
Unlike some other high-redshift BCGs \citep[e.g.][]{webb_star_2015,bonaventura_red_2017,cooke_stellar_2019}, ID 526 does not exhibit any signs of recent star formation. Its age distribution (see Figure \ref{fig_very_old_members}) is typical of a very old galaxy, albeit with slightly more probability toward 1 to 2 Gyrs than others, and its short characteristic time is inconsistent with the presence on a small population of young stars.

\subsection{ID 726}\label{ssec_726}
\begin{figure}%[h!]
\centering
\includegraphics[width=8cm]{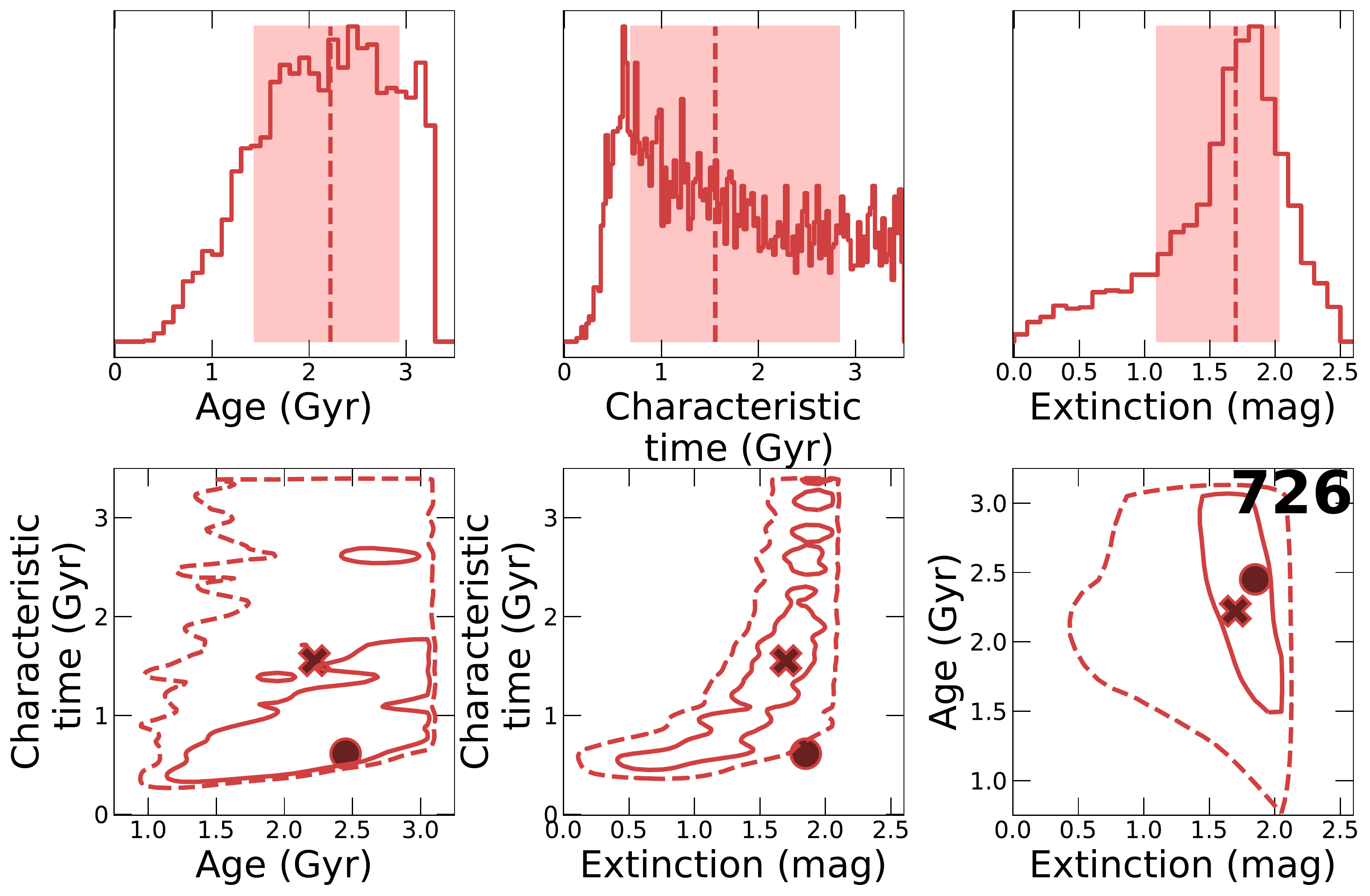}

\caption{\textit{Top}: Age, $\tau$ and dust extinction posterior distribution for ID 726, a dust-rich galaxy. \textit{Bottom}: Age-$\tau$, $A_v$-$\tau$ and Age-$A_v$, degeneracy plots for ID 726. Medians, modes and 1 and 2$\sigma$ contours are presented following the same convention as in Figure \ref{fig_very_old_members}.}
\label{fig_dusty_members}
\end{figure}

ID 726 (presented in Figure \ref{fig_dusty_members}) is the only reliable fit that does not enter into any of the four main age categories. Although its age distribution corresponds to a very old member, its characteristic time is poorly constrained and its dust content is markedly higher than in any other fit. Together with its irregular shape (see the top panel of Figure \ref{fig_XLSSC122_members}), this might suggest that ID 726 has undergone (or is still experiencing) an highly obscured star-forming episode. %ADD SOMETHING ABOUT ID 145

\subsection{ID 522}\label{ssec_522}
ID 522 characteristics place it between the young and the star-forming members: it is younger and bluer than any other evolved member (see Figure \ref{fig_young_members} and the bottom panel of Figure \ref{fig_XLSSC122_members}). However, unlike the other blue cloud members its characteristic time is very short and well-constrained, which could mean that ID 522 recently transitioned from star-forming to quenched. We thus classified it as a young members.

\section{Examples of full parameters distributions}\label{sec_corner_plots}
\begin{figure*}%[!h]
\centering
%~\qquad
\centering
\includegraphics[width=15cm]{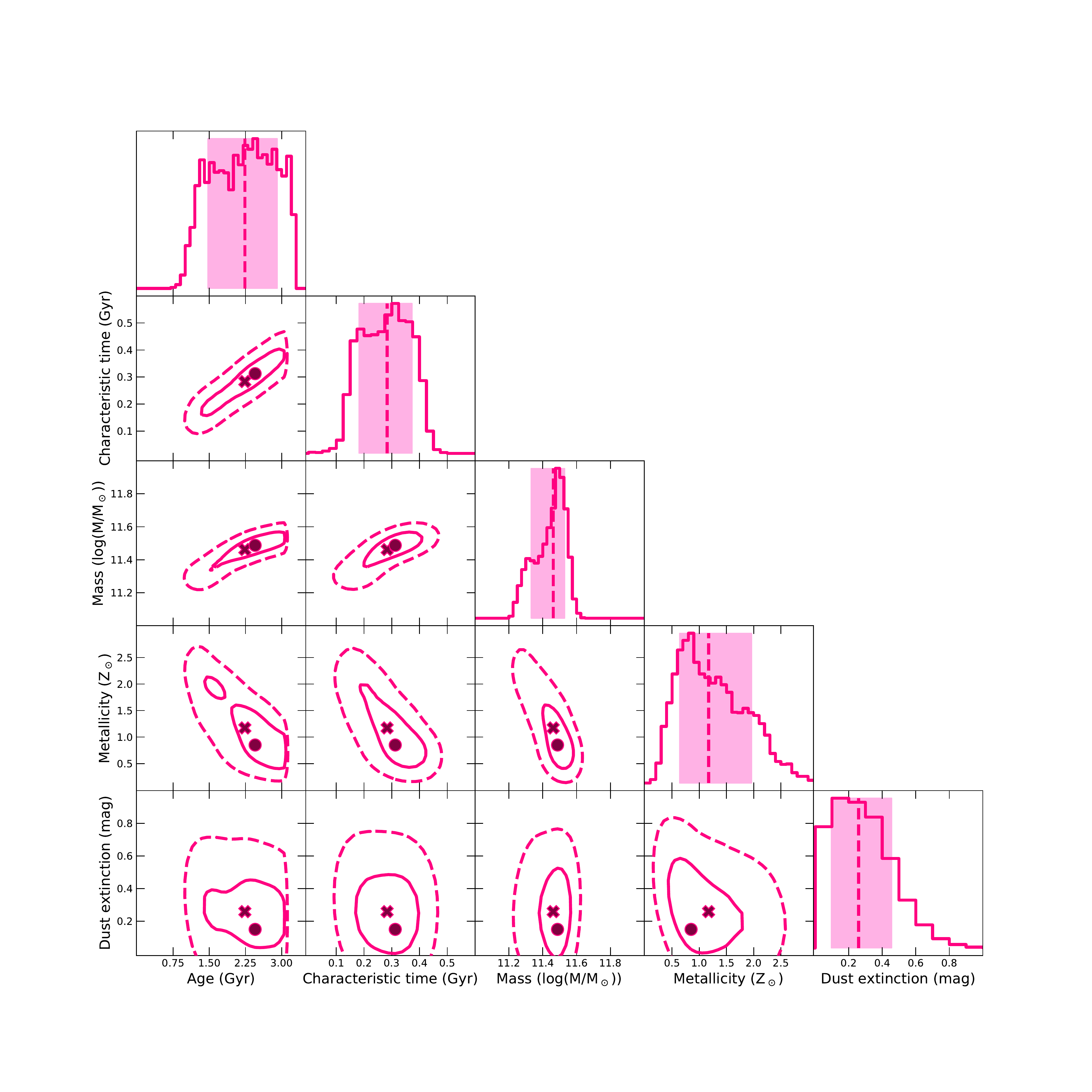}
\centering
\includegraphics[width=10cm]{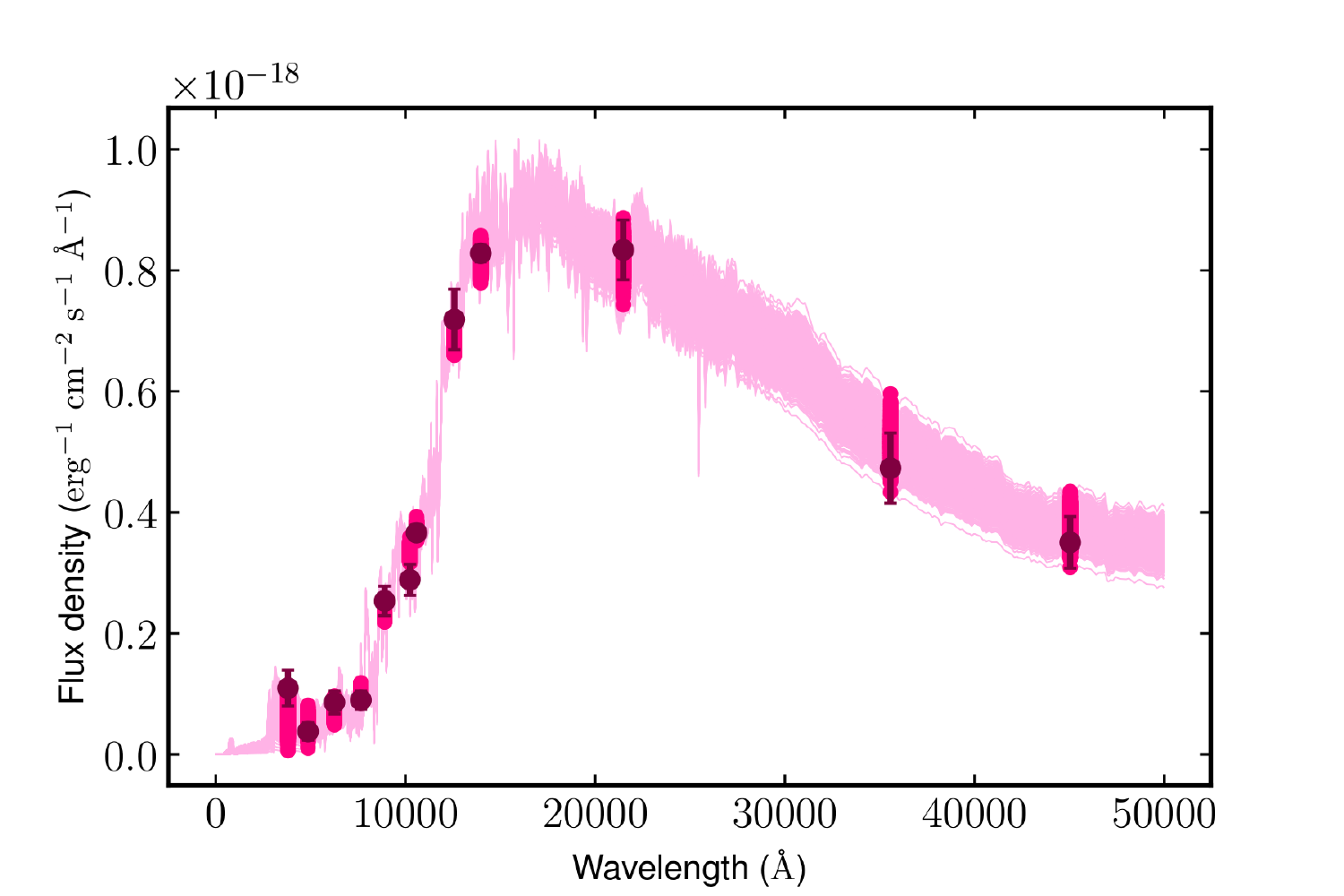}
\caption{\textit{Top}: A corner plot showing the results of the BCG fits, i.e. every parameter distribution and degeneracies. On the panels showing posterior distributions, the distances between the 16th and 84th percentiles (i.e. the 1$\sigma$ confidence region) are coloured and the medians are highlighted by dashed lines. On the panels illustrating degeneracies, the 1 and 2$\sigma$ contours are shown. Modes and medians are respectively denoted by dots and Xs. \textit{Bottom}: Comparison between the BCG flux density measurements ($f_\lambda$) and their best fits. The shaded region corresponds to the fits between the 16th and 84th percentiles and the highlighted spots correspond to the theoretical flux densities measurements associated with those fits. Observed flux densities are represented as dark dots.} %\textcolor{red}{The mass posteriors displayed here represent the total stellar masses formed, and are thus more elevated than the stellar masses presented in Table \ref{table_members}.}}
\label{fig_corner_526}
\end{figure*}

\begin{figure*}%[!h]
\centering
%~\qquad
\centering
\includegraphics[width=15cm]{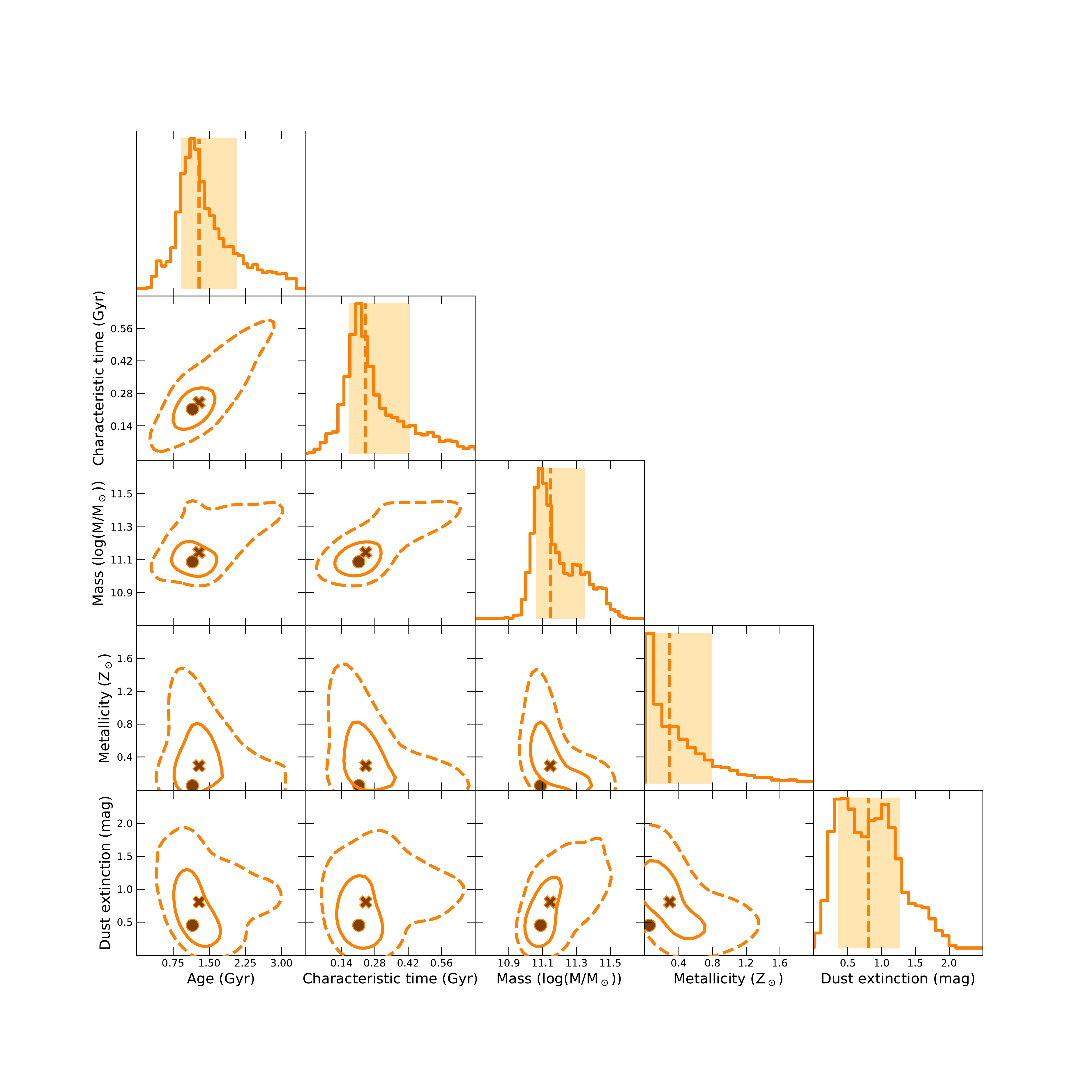}
\centering
\includegraphics[width=10cm]{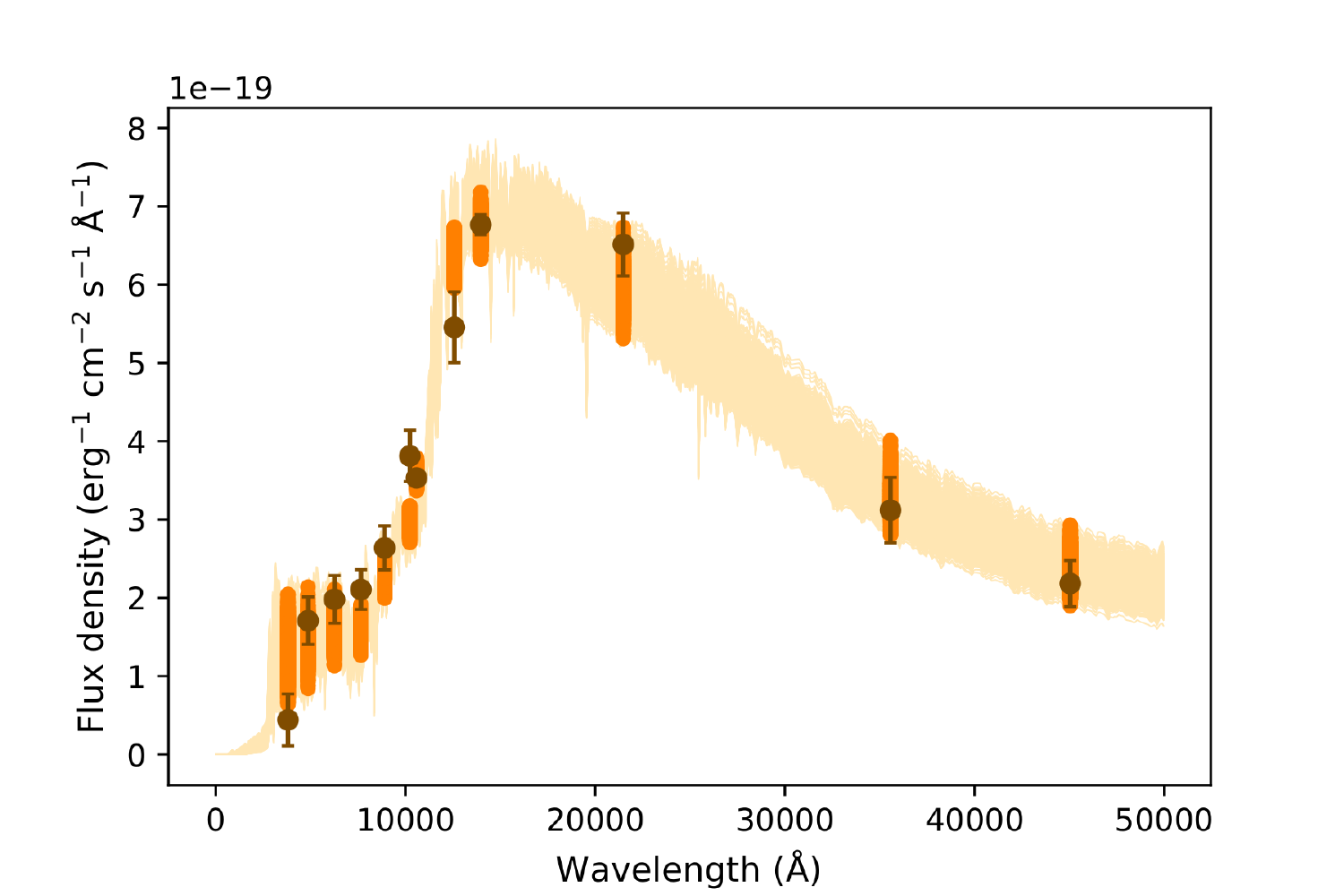}
\caption{\textit{Top}: ID 451 corner plot. \textit{Bottom}: ID 451 best fits. See the previous Figure for an explanation of the symbols.}
\label{fig_corner_451}
\end{figure*}

\begin{figure*}%[!h]
\centering
%~\qquad
\centering
\includegraphics[width=15cm]{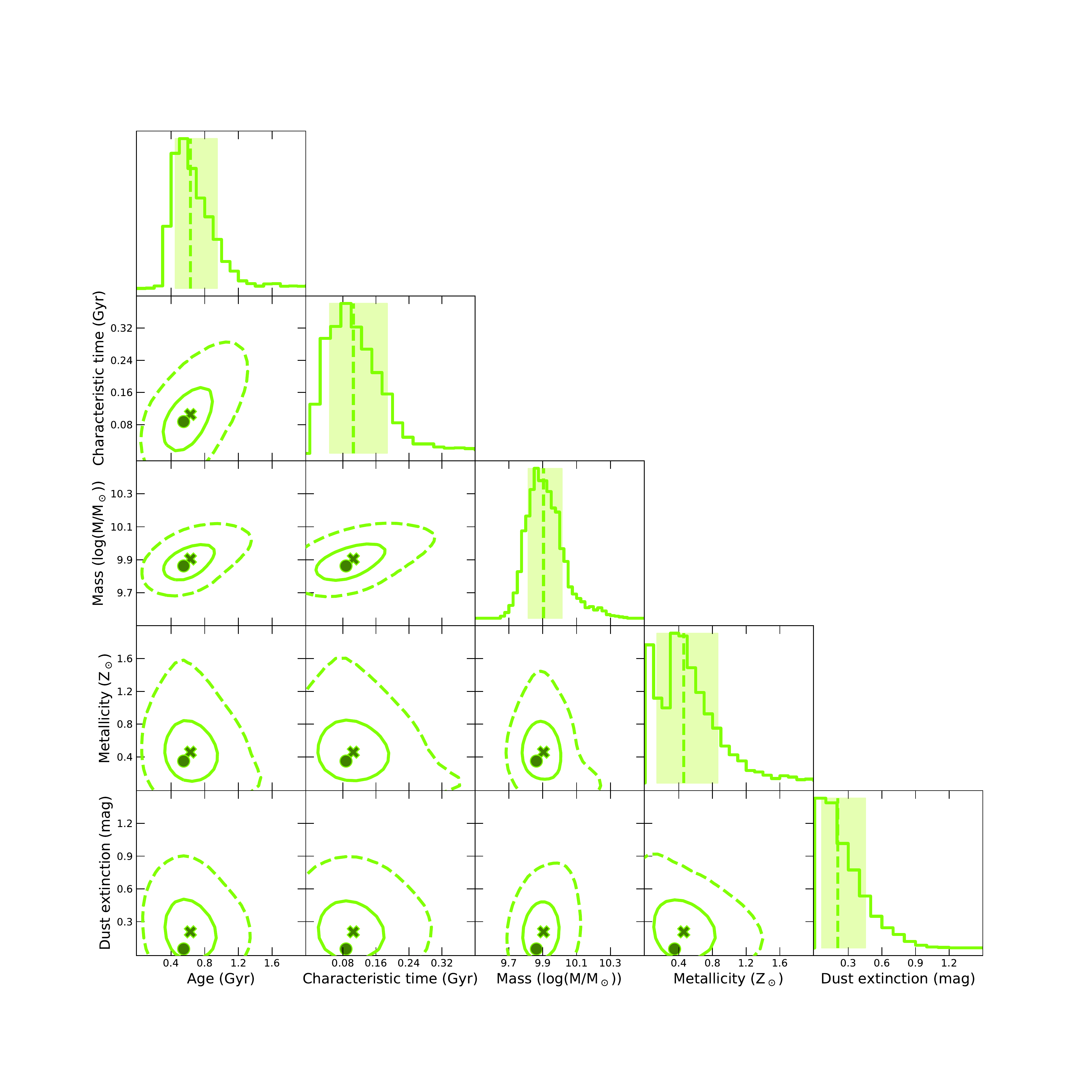}
\centering
\includegraphics[width=10cm]{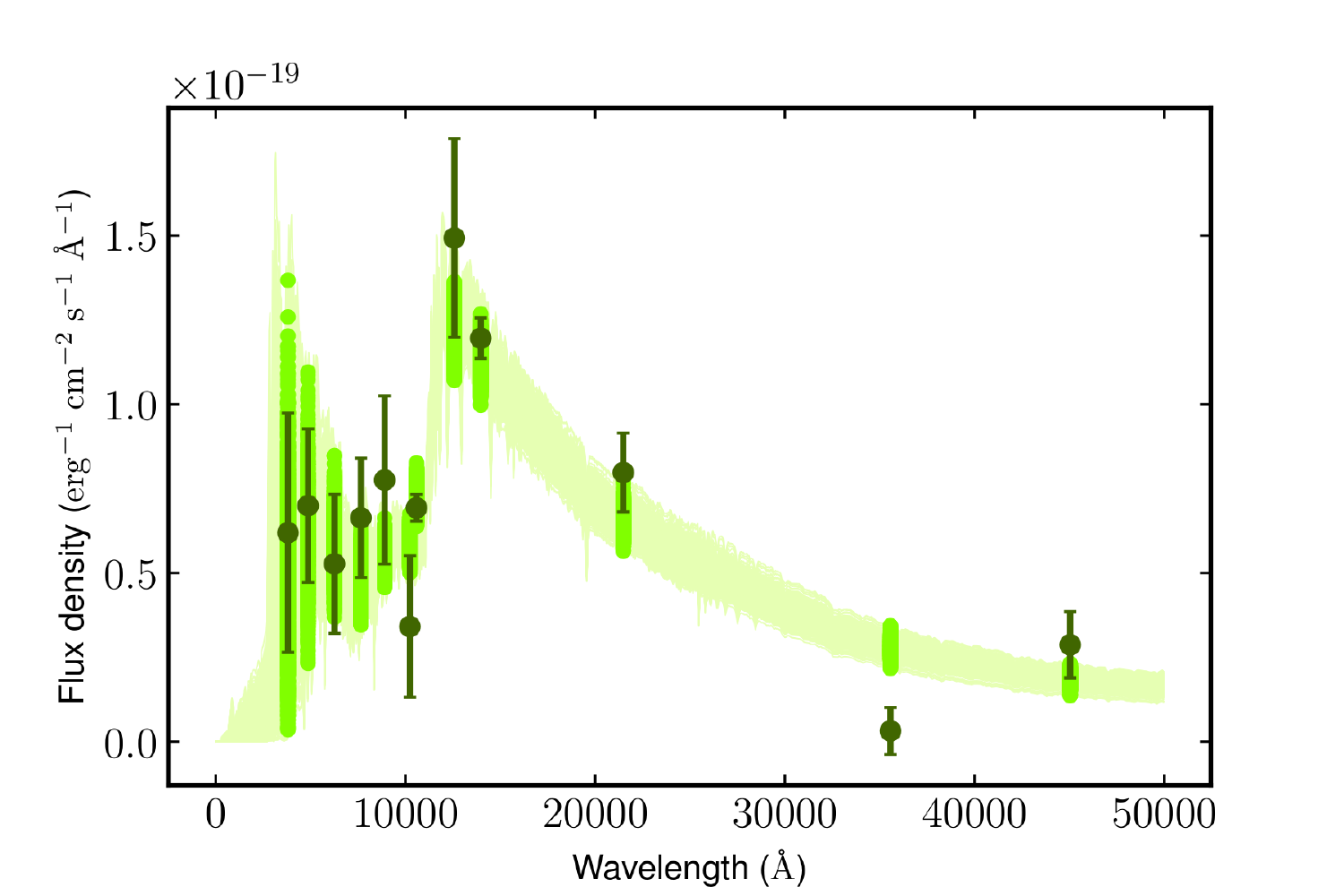}
\caption{\textit{Top}: ID 806 corner plot. \textit{Bottom}: ID 806 best fits. See Figure \ref{fig_corner_526} for an explanation of the symbols.}
\label{fig_corner_806}
\end{figure*}

\begin{figure*}%[!h]
\centering
\centering
\includegraphics[width=15cm]{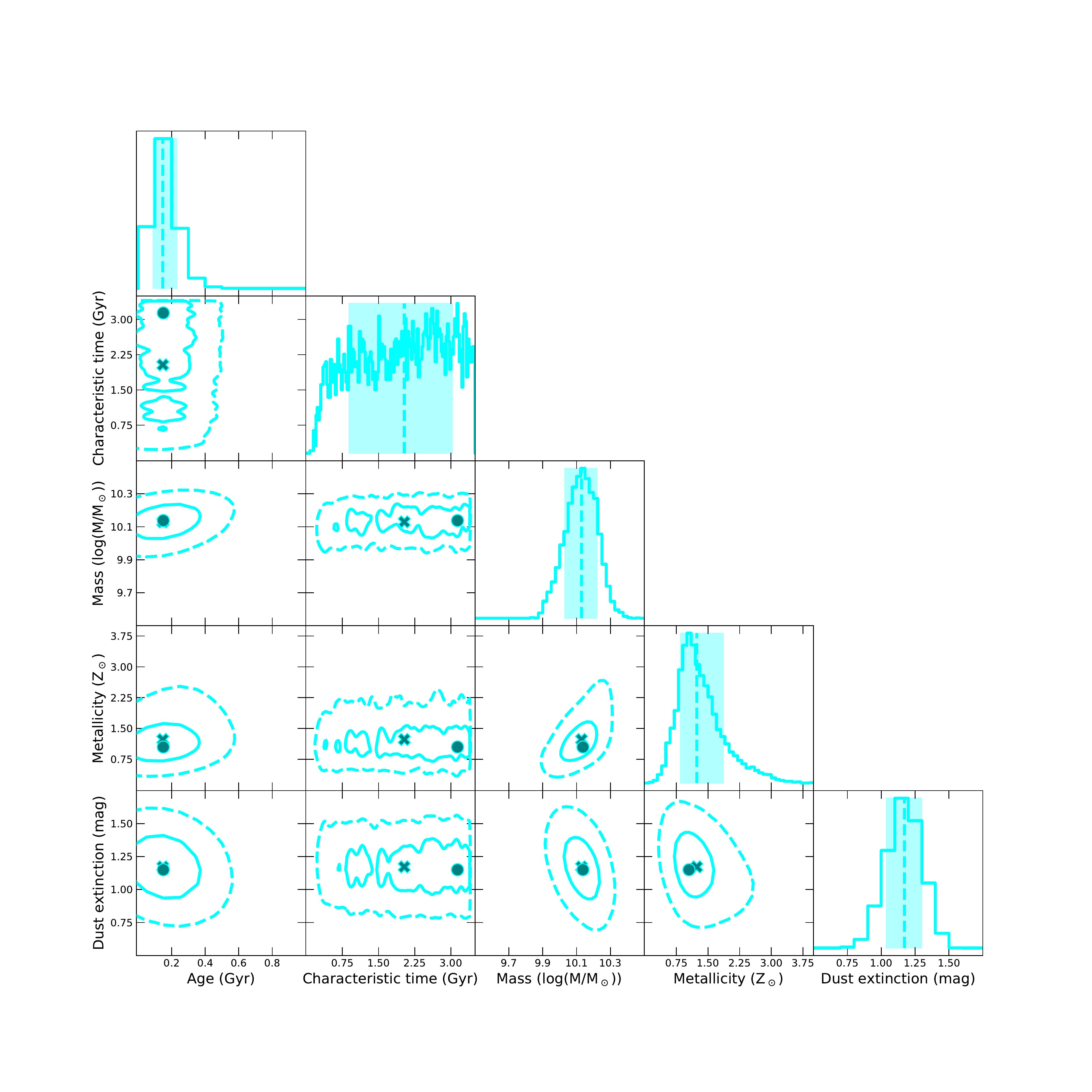}
\centering
\includegraphics[width=10cm]{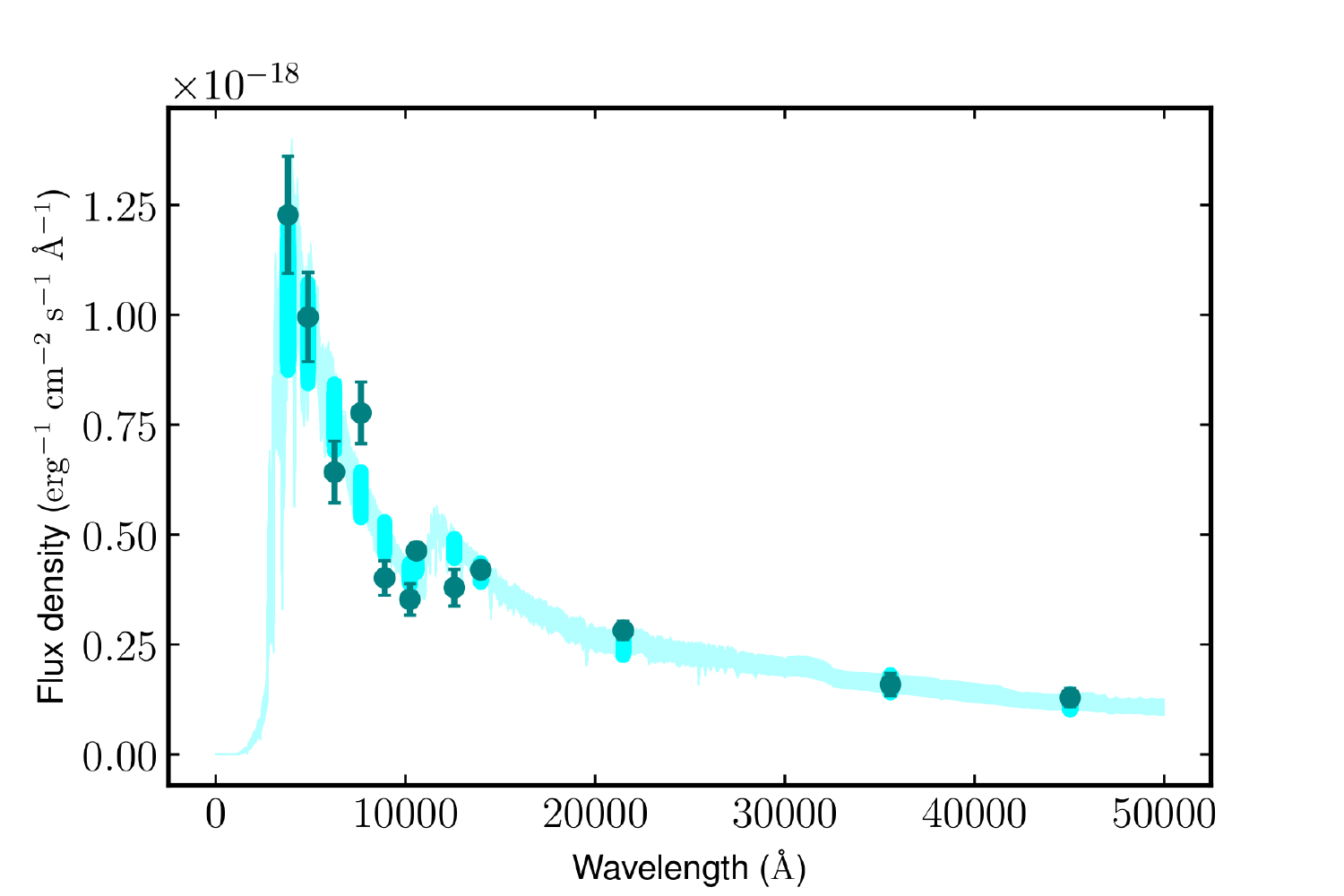}
\caption{\textit{Top}: ID 917 corner plot. \textit{Bottom}: ID 917 best fits. See Figure \ref{fig_corner_526} for an explanation of the symbols.}
\label{fig_corner_917}
\end{figure*}

Figures \ref{fig_corner_526} to \ref{fig_corner_917} present the full results of the fits for four XLSSC 122 members: the BCG, ID 451, ID 806 (members of the red sequence; Figure \ref{fig_corner_526}, \ref{fig_corner_451} and \ref{fig_corner_806} respectively) and ID 917 (in the blue cloud; \ref{fig_corner_917}). Each column shows the impact of one free parameter on the fits: the top panel displays the posterior distribution, and the subsequent panels show the degeneracies between this parameter and the others.
%Figures \ref{fig_corner_526} and \ref{fig_corner_917} present the full results of the fits for two XLSSC 122 members: the BCG (a member of the red sequence; Figure \ref{fig_corner_526}) and ID 917 (in the blue cloud; \ref{fig_corner_917}). Each column shows the impact of one free parameter on the fits: the top panel displays the posterior distribution, and the subsequent panels show the degeneracies between this parameter and the others.

\section{Mock photometry plots}\label{sec_mock_photometry_plot}

\begin{figure*}%[!h]
\centering
\includegraphics[width=15cm]{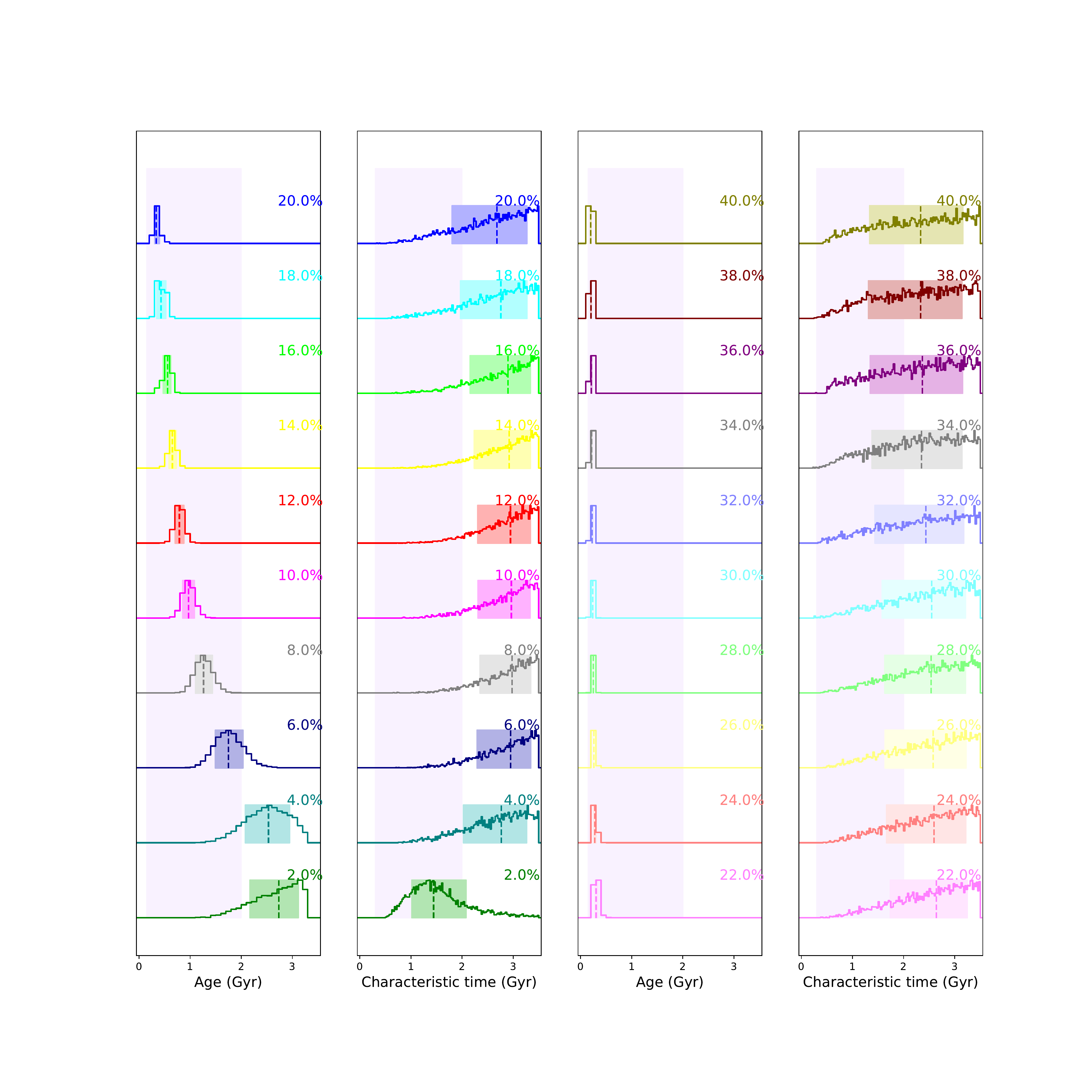}
\caption{The age and characteristic time distributions obtained by fitting the mock photometry presented in Section \ref{ssec_mock_photometry}. The percentages on top of each distribution indicate how much of the stellar mass was produced by the ongoing episode of star formation. Dashed lines indicate medians. The coloured regions highlight the 1$\sigma$ intervals associated with each distribution and the pale lilac shades indicate the difference between the parameters used to create the old and young components of the mock photometry.}
\label{fig_mock_hybrid_SED}
\end{figure*}

Figure \ref{fig_mock_hybrid_SED} shows the results of the mock photometry tests presented in Section \ref{ssec_mock_photometry}: simulated galaxies containing an evolved and a young stellar populations. The percentage of the simulated galaxies masses corresponding to young stars is indicated on the top of each plot. Even a small percentage of young stars has a significant impact on the characteristic time distribution, suggesting that the evolved members in our sample are unlikely to experience a significant amount of star formation.

\section{The cluster assembly history with a delayed $\tau$-model}\label{sec_delayed_assembly_history}

\begin{figure*}
\centering
\includegraphics[width=16.5cm]{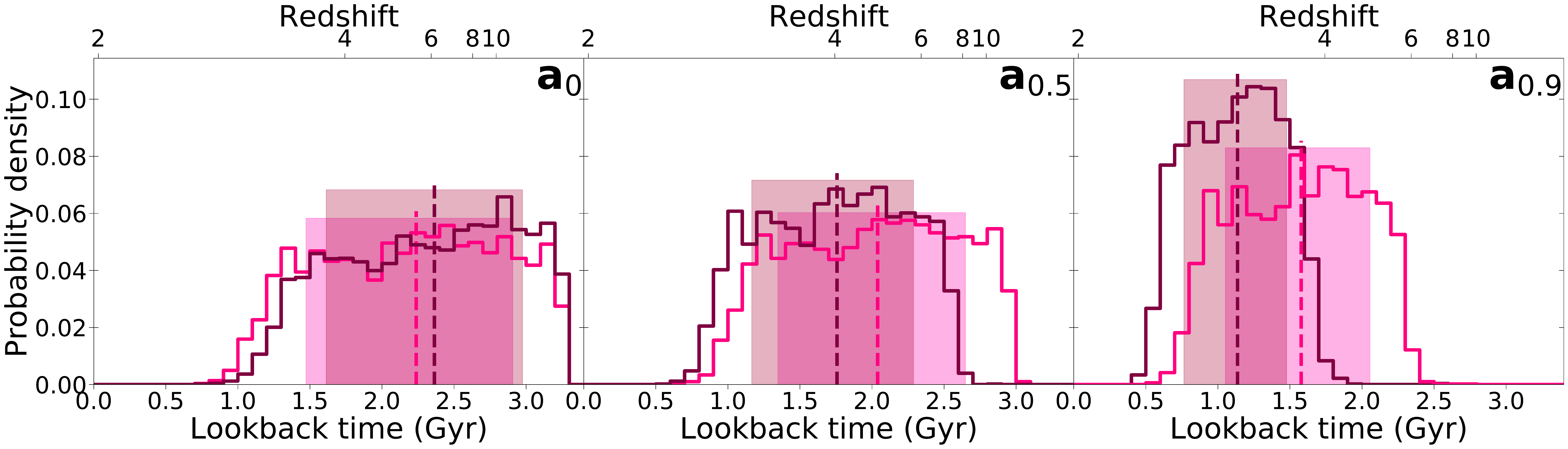} 
\caption{\textit{Left}: Comparison between the $a_0$ posteriors (i.e. the age of the oldest stars) of the simple $\tau$-model, in pink, and the delayed $\tau$-model, in burgundy, for the BCG. The dashed lines are the distribution medians and the shaded regions correspond to the intervals between the 16th and 84th percentiles. \textit{Centre}: Comparison between the distributions corresponding to the time-scale at which the BCG formed 50 per cent of its stellar mass (i.e. $a_{0.5}$) for the simple and delayed $\tau$-models. \textit{Right}: Comparizon between the $a_{0.9}$ distributions of the simple and delayed $\tau$-models.}
\label{fig_comparison_age_posteriors_526}
\end{figure*}

For the delayed $\tau$-model, the equivalent of Equation \ref{eq_a_X} is:
\begin{equation}\label{eq_a_X_delayed}
a_X=a_0+\tau+W_{-1}\left( \frac{-1}{\tau e} ((1-X)\tau+X(a_0+\tau)e^{-a_0/\tau})\right),
\end{equation}

\noindent where $W_{-1}$ is one of the branches of the Lambert W-function. Figure \ref{fig_comparison_age_posteriors_526} displays comparisons between the $a_0$,  $a_{0.5}$ and $a_{0.9}$ posteriors of the simple and delayed $\tau$-models, for the BCG. The $a_{0}$ posteriors, corresponding to the age of the oldest stars, are very similar. However, the different star formation history, combined with a slightly shorter characteristic time generate an $a_{0.9}$ distribution slightly younger and more peaked than the one corresponding to the regular $\tau$-model. Tables \ref{table_assembly_formation_time_delayed}, \ref{table_assembly_50_time_delayed}, and \ref{table_assembly_90_time_delayed} present our assembly predictions for XLSSC 122-like halo, assuming a delayed $\tau$-model.

\begin{table*}%[!ht]
\caption{Cumulative fractions of XLSSC 122 haloes that are $<$10\% assembled, $<$20\% assembled, etc., integrated over the age distributions of the oldest members, assuming a delayed $\tau$-model.}
%\caption{Cumulative probabilities of assembly of XLSSC 122, integrated over the age distributions of the oldest members.}
\label{table_assembly_formation_time_delayed}
\centering
%\setlength{\tabcolsep}{2pt}
%%\tablenum{1}
\begin{tabular}{c c c c c}
\hline
ID & $<$10\% & $<$20\% & $<$30\% & $<$40\%\\
 %\\
 \hline
526 & 0.80 & 0.91 & 0.96 & 0.98\\
657 & 0.74 & 0.85 & 0.92 & 0.96\\
295 & 0.79 & 0.88 & 0.93 & 0.96\\
606 & 0.76 & 0.87 & 0.92 & 0.96\\
734 & 0.62 & 0.77 & 0.87 & 0.92\\
845 & 0.82 & 0.89 & 0.93 & 0.96\\
493 & 0.73 & 0.85 & 0.91 & 0.95\\
730 & 0.74 & 0.84 & 0.90 & 0.94\\
average & 0.75 & 0.86 & 0.92 & 0.95\\
\hline
\end{tabular}
\end{table*}

\begin{table*}%[!h]
\caption{Cumulative fraction of XLSSC 122-like haloes that are partially assembled, integrated over the time distributions corresponding to the formation of 50\% of the stellar masses of the oldest members, assuming a delayed $\tau$-model.}
\label{table_assembly_50_time_delayed}
\centering
%\setlength{\tabcolsep}{2pt}
%%\tablenum{1}
\begin{tabular}{c c c c c c}
\hline
ID & $<$10\% & $<$20\% & $<$30\% & $<$40\% & $<$50\%\\
 %\\
 \hline
526 & 0.67 & 0.83 & 0.91 & 0.96 & 0.98\\
657 & 0.63 & 0.77 & 0.87 & 0.92 & 0.96\\
295 & 0.67 & 0.80 & 0.88 & 0.93 & 0.95\\
606 & 0.55 & 0.72 & 0.82 & 0.89 & 0.93\\
734 & 0.54 & 0.71 & 0.82 & 0.89 & 0.93\\
845 & 0.74 & 0.85 & 0.90 & 0.94 & 0.96\\
493 & 0.65 & 0.79 & 0.87 & 0.92 & 0.95\\
730 & 0.54 & 0.70 & 0.81 & 0.87 & 0.91\\
average & 0.62 & 0.77 & 0.86 & 0.92 & 0.95\\
\hline
\end{tabular}
\end{table*}

\begin{table*}%[!h]
\caption{Cumulative fraction of XLSSC 122-like haloes that are partially assembled, integrated over the time distributions corresponding to the formation of 90\% of the stellar masses of the oldest members, assuming a delayed $\tau$-model.}
\label{table_assembly_90_time_delayed}
\centering
%\setlength{\tabcolsep}{2pt}
%%\tablenum{1}
\begin{tabular}{c c c c c c c c}
\hline
ID & $<$10\% & $<$20\% & $<$30\% & $<$40\% & $<$50\% & $<$60\% & $<$70\%\\
 %\\
 \hline
526 & 0.34 & 0.59 & 0.76 & 0.86 & 0.92 & 0.95 & 0.97\\
657 & 0.36 & 0.58 & 0.73 & 0.83 & 0.89 & 0.93 & 0.96\\
295 & 0.33 & 0.57 & 0.73 & 0.82 & 0.88 & 0.92 & 0.95\\
606 & 0.10 & 0.27 & 0.46 & 0.61 & 0.72 & 0.79 & 0.86\\
734 & 0.40 & 0.59 & 0.73 & 0.83 & 0.89 & 0.93 & 0.96\\
845 & 0.46 & 0.68 & 0.81 & 0.88 & 0.92 & 0.95 & 0.97\\
493 & 0.48 & 0.66 & 0.78 & 0.86 & 0.91 & 0.94 & 0.96\\
730 & 0.16 & 0.34 & 0.51 & 0.63 & 0.72 & 0.79 & 0.85\\
average & 0.33 & 0.54 & 0.69 & 0.79 & 0.86 & 0.90 & 0.94\\
\hline
\end{tabular}
\end{table*}

%
%%%%%%%%%%%%%%%%%%%%%%%%%%%%%%%%%%%%%%%%%%%%%%%%%%%

% Don't change these lines
\bsp	% typesetting comment
\label{lastpage}
\end{document}